\documentclass[aps,prl,reprint,showpacs,amsfonts,amsmath,floatfix,superscriptaddress,footinbib]{revtex4-2}

\usepackage[svgnames]{xcolor}

 \usepackage[colorlinks=true,        
            allcolors = black,  
            citecolor=DarkBlue, 
            linkcolor=DarkBlue,
            urlcolor=DarkBlue
        ]{hyperref}  
\usepackage{amsmath, amsfonts, amssymb, amsthm, bbm}
\usepackage{dsfont}
\usepackage{bm} 
\usepackage{mathtools}
\usepackage{times}
\usepackage{physics}
\usepackage{graphicx}
\usepackage{float}
\usepackage{enumitem}
\usepackage{dcolumn}

\usepackage{diagbox}

\newcommand{\FF}{\mathbb{F}} 
\newcommand{\EE}{\mathbb{E}} 

\newcommand{\umag}{\mathcal{G}} 
\newcommand{\logneg}{\mathcal{W}} 
\newcommand{\cellnegint}{\mathcal{N}_0} 
\newcommand{\cellneg}{\mathcal{N}} 

\newcommand{\ddqn}{\dd^n {\bm{q}}}
\newcommand{\ddpn}{\dd^n {\bm{p}}}
\newcommand{\ddrn}{\dd^n {\bm{r}}}
\newcommand{\qn}{\bm{q}}
\newcommand{\pn}{\bm{p}}

\newcommand{\rvline}{\hspace*{-\arraycolsep}\vline\hspace*{-\arraycolsep}}
\allowdisplaybreaks

\usepackage{subfiles}

\begin{document}
\title{Quantifying Qubit Magic Resource with Gottesman-Kitaev-Preskill Encoding}

\author{Oliver Hahn}
\email{hahno@chalmers.se}
\affiliation{Department of Microtechnology and Nanoscience (MC2), Chalmers University of Technology, SE-412 96 G\"{o}teborg, Sweden}

\author{Alessandro Ferraro}

\affiliation{Centre for Theoretical Atomic, Molecular and Optical Physics, Queen's University Belfast, Belfast BT7 1NN, United Kingdom}

\author{Lina Hultquist}
\affiliation{Department of Microtechnology and Nanoscience (MC2), Chalmers University of Technology, SE-412 96 G\"{o}teborg, Sweden}

\author{Giulia Ferrini}
\affiliation{Department of Microtechnology and Nanoscience (MC2), Chalmers University of Technology, SE-412 96 G\"{o}teborg, Sweden}

\author{ Laura Garc\'{\i}a-\'{A}lvarez}
\affiliation{Department of Microtechnology and Nanoscience (MC2), Chalmers University of Technology, SE-412 96 G\"{o}teborg, Sweden}


\begin{abstract}
Quantum resource theories are a powerful framework to characterize and quantify relevant quantum phenomena and identify processes that optimize their use for different tasks. Here, we define a resource measure for magic, the sought-after property in most fault-tolerant quantum computers. In contrast to previous literature, our formulation is based on bosonic codes, well-studied tools in continuous-variable quantum computation. Particularly, we use the Gottesman-Kitaev-Preskill code to represent multi-qubit states and consider the resource theory for the Wigner negativity. 
Our techniques are useful to find resource lower bounds for different applications as state conversion and gate synthesis. The analytical expression of our magic measure allows us to extend current analysis limited to small dimensions, easily addressing systems of up to 12 qubits.
\end{abstract}

\maketitle

Identifying and quantifying which properties of quantum mechanics, or {\it resources}, are responsible for the predicted advantage of quantum computers over classical ones is the object of intense theoretical and experimental effort. Within the field of resource theories~\cite{brandao2015reversible, coecke2016mathematical, grudka2014quantifying,  horodecki2013quantumness, napoli2016robustness, stahlke2014quantum, vidal1999robustness, veitch2014the-resource, albarelli2018resource, takagi2018convex}, the quantification of magic for fault-tolerant quantum computation occupies a prominent role. The leading architectures of fault-tolerant quantum computers are based on stabilizer codes~\cite{fowler2012surface}. In this approach, certain operations are easy to implement and constitute a non-universal~\cite{eastin2009restrictions} set that is fault-tolerant by having a transversal implementation ---such that these operations do not propagate errors within a code block~\cite{gottesman1998theory}. This restricted set of easy or free operations called stabilizer operations includes Clifford gates, preparation of stabilizer states, and computational basis measurements. Stabilizer operations alone cannot provide quantum computational advantage, as the calculations can be efficiently classically simulated~\cite{gottesman1999the-heisenberg}. To unlock universal quantum computation, one needs to include difficult or resourceful operations
---for example, through magic state injection---
which naturally leads to the so-called magic state model~\cite{gottesman1999demonstrating}. However, the preparation of high-quality magic states typically involves costly procedures such as magic state distillation, which makes it desirable to optimise 
the number of magic states used for a given quantum computation --- although alternative routes to magic state distillation exist~\cite{brown2020a-fault-tolerant}.

The need for optimizing non-stabilizer resources has driven the area of resource theory of magic and therefore the definition of several magic measures for both qudit ---$d$-dimensional--- and qubit systems. Early works focused on resource theories of magic for odd-dimensional qudits~\cite{koukoulekidis2021constraints}, relying on a well-defined discrete Wigner function and its negativity~\cite{veitch2014the-resource} , and its extensions to infinite dimensions~\cite{,albarelli2018resource, takagi2018convex}. The more defying case of multi-qubit systems ---for which the definition of the discrete Wigner function remains challenging~\cite{delfosse2015wigner,raussendorf2017contextuality,kocia2017discrete,raussendorf2020phase-space-simulation}--- has undergone substantial progress in recent years with the development of several magic measures. Among these, the  relative entropy of Magic~\cite{veitch2014the-resource}, the Robustness of Magic~\cite{howard2017application}, the dyadic negativity, the mixed state extent, and the generalized robustness~\cite{seddon2021quantifying} have been defined for general density matrices, while the stabilizer rank or extent~\cite{bravyi2016improved, bravyi2016trading, bravyi2019simulation}, the stabilizer nullity, and the dyadic monotone~\cite{beverland2020lower} only account for the magic content of pure states.  
The availability of these measures has been essential to find (in few cases optimal) lower bounds in magic state distillation schemes 
and in non-Clifford unitary synthesis~\cite{howard2017application, beverland2020lower, seddon2021quantifying}. Moreover, magic monotones have inspired classical simulators of quantum computing architectures~\cite{bravyi2016improved, bravyi2016trading, bravyi2019simulation, seddon2021quantifying}. 
However, in general, it is impractical to compute these measures for large numbers of qubits ($\gtrsim 5$), and several lower bounds in the literature apply to distinct scenarios in which not always all easy operations are considered free ---for instance, the measures do not always account for measurements and classical feed-forward as free operations. Therefore, it is desirable to provide new measures that are practically computable for larger number of qubits, and combine different quantifiers of magic to find tighter bounds applicable to general scenarios and to identify how quantum computations can be optimized. 

Here, we develop a new magic measure for multi-qubit pure states by borrowing tools and notions from continuous-variable (CV) systems in the context of bosonic codes. Moreover, we connect this independently developed magic measure with the st-norm~\cite{campbell2011catalysis, howard2017application}, and with the stabiliser R\'enyi entropy~\cite{leone2022stabilizer}, measured recently on a quantum processor~\cite{oliviero2022measuring}. The CV framework provides a way to upgrade the st-norm status to a fully-fledged measure and brings new insight on its properties and tools for its computation. We exploit the encoding of discrete-variable systems into the infinite-dimensional Hilbert space of CV systems provided by bosonic codes, the main alternative to conventional error-correction codes with growing theoretical and experimental efforts~\cite{Terhal_2020,joshi2021quantum, CAI202150, grimsmo2021quantum}. In contrast to all previous magic monotones, the CV mathematical formulation of our monotone allows for transferring results from CV quantum computation to magic quantification. 
Specifically, we map qubit systems into an infinite-dimensional Hilbert space via the Gottesman-Kitaev-Preskill (GKP) encoding~\cite{gottesman2001encoding} and then derive an expression for their Wigner logarithmic negativity (WLN)~\cite{kenfack2004negativity,albarelli2018resource, takagi2018convex}. Using this expression, we develop a new magic measure that we call GKP Magic. The Wigner function is well-defined for CV systems and allows us to evaluate the GKP Magic for significantly larger systems than previously known measures, easily reaching up to 12-qubit systems. The GKP Magic properties that we prove enable the analysis of the magic resource cost in the most general scenario, where probabilistic protocols are allowed, and measurements, auxiliary qubits, and classical feed-forward are free operations. In this context, we find analytical expressions of the GKP Magic for relevant building blocks of quantum algorithms yielding lower bounds for the corresponding $T$-count, a known indicator of the difficulty to implement fault-tolerant quantum circuits.

\paragraph*{Definition of the GKP Magic measure.---}
We consider a general $n$-qubit state in the $\hat{Z}$ eigenbasis, represented by the density operator
\begin{align}
\label{eq:generalrho}
    \hat{\rho}= \sum_{\bm{u},\bm{v} \in \FF_2^n} \rho_{\bm{u},\bm{v}}\ket{\bm{u}}\bra{\bm{v}},
\end{align}
with $\ket{\bm{u}} = \ket{u_1 \dots u_n}$ and $\ket{\bm{v}} = \ket{v_1 \dots v_n}$ a tensor product of the single-qubit states in the computational basis, $\FF_2^n$ the $n$-dimensional binary linear space, and $\rho_{\bm{u},\bm{v}}$ complex coefficients. With the GKP encoding in square lattices, the code words $\ket{u_i}$ ---with $u_i\in\{0,1\}$--- correspond to the infinite superpositions of position $\hat{q}$ eigenstates
\begin{align}
    \ket{u_i} = \sum_{s_i= -\infty}^\infty \ket{x_i = \sqrt{\pi}(u_i +2 s_i)}_{\hat{q}}.
\end{align}
Then, the Wigner function associated to the GKP-encoded density operator of Eq.~(\ref{eq:generalrho}) is given by
\begin{align}
\label{eq:n_qubit_W}
    &W_{\hat{\rho}} (\qn,\pn) \equiv \frac{1}{(2\pi)^n}\int_{-\infty}^\infty \dd^n\bm{x} e^{i\pn \bm{x}} \bra{\qn+\frac{\bm{x}}{2}} \hat{\rho}\ket{\qn-\frac{\bm{x}}{2}}_{\hat{q}} \nonumber\\
    & \qquad =\frac{1}{(2\pi)^n} \sum_{\bm{u},\bm{v} \in \FF_2^n} \rho_{\bm{u},\bm{v}} \prod_{i=1}^n \Bigg[ \sum_{s_i,t_i}(-1)^{\frac{s_i}{2}(u_i-v_i-2t_i)}  \nonumber \\
    & \qquad \times \delta\qty(p_i-\frac{\sqrt{\pi}}{2}  s_i) \delta\qty(q_i-\frac{\sqrt{\pi}}{2}(2t_i+u_i+v_i)) \Bigg],
\end{align}
and it constitutes the foundation of the GKP Magic measure. 

The negativity of the Wigner function is a necessary condition for achieving exponential speed-up in continuous-variable quantum computing architectures~\cite{mari2012positive}.
A measure of this resource is the WLN, defined as~\cite{albarelli2018resource, takagi2018convex} 
\begin{align}
\label{eq:Wigner_log_neg}
    \logneg(\hat{\sigma}) = \log_2 \qty( \int_{-\infty}^\infty \ddqn \ddpn  |W_{\hat{\sigma}}(\qn,\pn)|),
\end{align}
with $\ddqn$ and $\ddpn$ the $n$-dimensional volume differentials corresponding to $\qn$ and $\pn$, and
$\hat{\sigma}$ a bosonic Hermitian operator.

Firstly, we compute the negativity of the Wigner function given in Eq.~(\ref{eq:n_qubit_W}). Since the WLN of an ideal non-normalized GKP codeword is infinite, we consider the periodicity of the Wigner function and restrict the computation to the lattice unit cell to obtain a finite value. We reduce the integration domain in Eq.~(\ref{eq:Wigner_log_neg}) to a hypercube in phase space, with the domain $\mathcal{C}$ being  $q_i\in[0,2\sqrt{\pi}) ,p_i\in[0,2\sqrt{\pi})$.
Therefore, we define the WLN of one unit cell as
\begin{align}
\label{eq:sum_comp}
    \logneg_{C}(\hat{\rho}) =  \log_2 \qty( \int_\mathcal{C} \ddqn \ddpn  |W_{\hat{\rho}}(\qn,\pn)|).
\end{align}
The integral over the absolute values of the Wigner function can be evaluated and is obtained as
\begin{align}
\label{eq:sum_w}
    \int_\mathcal{C} \ddqn \ddpn  |W_{\hat{\rho}}(\qn,\pn)| 
    = \frac{1}{\sqrt{\pi}^n} \sum_{\bm{i},\bm{j} \in \FF_2^n}  \abs{ \sum_{\bm{k} \in \FF_2^n} \qty(-1)^{\bm{i}\cdot \bm{k} }  \rho_{\bm{k},\bm{k}+\bm{j}} },
\end{align}
with $\bm{i}\cdot \bm{k} =\sum_{j=1}^n  i_j k_j \mod 2$ the standard binary inner product, and $\bm{k}+\bm{j}$ the bitwise sum $(\bm{k}+\bm{j})_i= k_i + j_i \mod 2 $.
(see Supplemental Material~{\cite[Sec. I]{supmat}}.)

Quantifying the cell Wigner negativity of a $n$-qubit GKP state in Eq.~(\ref{eq:sum_w}) allows us to define the GKP Magic $\umag(\ket{\psi})$, a new Magic monotone for pure states. This definition was first motivated by noticing that the cell WLN of encoded GKP states saturates to a constant value for encoded stabilizer states, while it is maximal for the $\ket{T}$ and $\ket{H}$ magic states~\cite{garcia-alvarez2021from, yamasaki2020cost-reduced}.
 We emphasize the generality of the previous Wigner function logarithmic negativity calculations by using $\rho$ to denote any mixed or pure state. For the sake of clarity, since we only demonstrate the monotonic properties of the GKP Magic for pure states $\rho=\ket{\psi}\bra{\psi}$, we stress the difference by denoting the quantum state as $\ket{\psi}$ from now on.
 
Crucially, we notice that GKP-encoded pure stabilizer states contain an inherent amount of WLN in one lattice cell. Therefore, we define our GKP Magic measure by subtracting the inherent cell negativity of $\qty(2/\sqrt{\pi})^n$ to enforce that $\umag(\ket{\psi_S}) =0$ for $\ket{\psi_S}$ a pure stabilizer state. 
We provide an explicit counterexample for mixed states in~{\cite[Sec. IV]{supmat}}.

In the case of pure states $\ket{\psi}=\sum_{\bm{i} \in \FF_2^n} c_{\bm{i}} \ket{\bm{i}}$, the GKP Magic measure is finally obtained as
\begin{align}
        \umag(\ket{\psi}) &\equiv  \log_2 \qty[\qty(\frac{\sqrt{\pi}}{2})^{n} \int_{\mathcal{C}} \ddqn \ddpn \; |W_{\ket{\psi}\bra{\psi}}(\qn,\pn)|) ]
        \nonumber \\
      &=  \log_2\qty( \sum_{\bm{i},\bm{j} \in \FF_2^n}  \abs{ \sum_{\bm{k} \in \FF_2^n} \frac{\qty(-1)^{\bm{i}\cdot \bm{k} }}{2^n}  c_{\bm{k}}^* c_{\bm{k}+\bm{j}} }) 
        \label{eq:MM}
\end{align}
where we made use of Eq.~(\ref{eq:sum_w}). 

The Wigner negativity in the argument of the logarithm is equivalent to the st-norm~{\cite[Sec. III]{supmat}}, initiallly regarded as a one-way magic witness. In turn, this also implies that it is equivalent to the stabiliser R\'enyi entropy~\cite{leone2022stabilizer}\cite[Sec. III]{supmat} for $\alpha=\frac{1}{2}$. These equivalences upgrade the st-norm to a fully fledged magic measure.

Using the properties of the WLN enables us to demonstrate the following properties for our GKP Magic measure $\umag$~\cite[Sec. II]{supmat}:
\begin{enumerate}[itemsep=-0.25em,label=(\roman*)]
    \item Invariance under Clifford unitaries $\hat{U}_C$: $\umag(\hat{U}_C\ket{\psi}) = \umag(\ket{\psi})  $ 
    \item Additivity: $\umag(\ket{\psi}_A\otimes\ket{\phi}_B) =\umag(\ket{\psi})+\umag(\ket{\phi}) $
    \item Faithfulness: $\umag(\ket{\psi_S}) =0$ iff $\ket{\psi_S}$ is a stabilizer state 
    \item Invariance under composition with stabilizer states: $\umag(\ket{\psi}\otimes\ket{\phi_S}) =\umag(\ket{\psi})$
    \item Non-increasing under measurement in the computational basis.
    \item Non-increasing under Clifford operations conditioned on the outcomes of computational-basis measurements.
\end{enumerate}

Using our newly defined magic measure, we compute the most magic states and unitaries~{\cite[Sec. V]{supmat}}.

\paragraph*{Distillation and gate synthesis.---}
Magic monotones play a central role in the leading approaches to fault-tolerant quantum computation, and have been used to bound the number of resourceful states for state conversion and 
gate synthesis~\cite{howard2017application,beverland2020lower}. Additionally, fundamental bounds have been found on the Gaussian conversion between GKP-encoded Hadamard eigenstates $\ket{H}$ and the logical GKP-state $\ket{0}$ in continuous-variable settings~\cite{yamasaki2020cost-reduced}. Using our GKP Magic measure, we can lower bound the number of copies of a given resource state needed to implement a desired target unitary or to produce certain state when non-unitary and probabilistic protocols are allowed.

Firstly, we address distillation protocols to extract a particular target state. We consider a stabilizer protocol~\cite{veitch2014the-resource} ---a set of Clifford unitaries, composition with stabilizer states, computational basis measurements, and Pauli operations conditioned on measurement outcomes--- that converts $k$ copies of $\ket{\psi}$ to $m$ copies of the target state $\ket{\phi}$. The GKP Magic does not increase with such stabilizer protocol, and therefore we can bound the number of input resource states by
\begin{align}
    k \geq  m \frac{\umag(\ket{\phi})}{\umag(\ket{\psi})},
\end{align}
where we have used the additive property of our measure, $\umag(\ket{\psi}^{\otimes k})=k\umag(\ket{\psi})$. We notice that this property also allows us to establish bounds even when catalyst states ---loaned magic states returned at the end of the protocol--- are allowed.

Moreover, we analyze probabilistic stabilizer protocols for distillation that convert $k$ copies of an $r$-qubit state $\ket{\psi}$ to $m$ copies of the $s$-qubit target state $\ket{\phi}$ with probability $p$. That is, we consider stabilizer protocols that can include post-selection upon specific measurement outcomes and operations entailing partial traces that can create mixed quantum states. Despite the GKP Magic monotone being only defined for pure states, its direct link with the WLN allows us to consider lower bounds of required resource states when intermediate mixed states are involved.
The system's WLN restricted to the code's unit cell is additive and does not increase on average with such probabilistic stabilizer protocol~{\cite[Sec. IV]{supmat}}, 
so that
\begin{align}
    k\logneg_{C}(\ket{\psi}) \geq  p m \logneg_{C}(\ket{\phi}).
\end{align}
We can establish a lower bound on the average number of copies $\EE[n]$ of $\ket{\psi}$ needed to distill $\ket{\phi}^{\otimes m}$ proportional to the ratio of the monotones. Since one must run the probabilistic protocol $1/p$ times to get a successful outcome, we require
\begin{align}
    \EE[n] = \frac{k}{p} \geq m \frac{\logneg_{C}(\ket{\phi})}{\logneg_{C}(\ket{\psi}) }.
\end{align}
Finally, for input and output pure states the WLN per cell is directly related with the GKP Magic as $\umag(\ket{\Psi})=\logneg_C(\ket{\Psi})-\log_2[\cellnegint(n)]$, where we have subtracted the corresponding intrinsic logarithmic negativity per cell of a pure $n$-qubit stabilizer state $\ket{\psi_S}$, given by $\cellnegint(n)=(2/\sqrt{\pi})^n$. Hence, we can rewrite the bound as
\begin{align}
    \EE[n] = \frac{k}{p} \geq m \frac{\umag(\ket{\phi})+\log_2[\cellnegint(s)] }{\umag(\ket{\psi})+\log_2[\cellnegint(r)] }.
\end{align}
This bound is strictly looser in the case of $p=1$~{\cite[Sec. IV]{supmat}}.

Besides characterizing distillation protocols, magic measures have been used to bound gate synthesis.  A quantum gate can be synthesized with purely unitary processes~\cite{forest2015exact} or, more generally, allowing auxiliary qubits, measurements and classical feed-forward~\cite{jones2013low, paetznick2013repeat, duclos2013distillation, wiebe2014quantum}. In the field of fault-tolerant quantum computation, gates of the third level of the Clifford hierarchy ---$C_3$ such that $C_{n+1}\equiv \{\hat{U}|\hat{U} \hat{P} \hat{U}^{\dagger} \subseteq C_{n},\forall \hat{P} \in C_1\}$~\cite{gottesman1999demonstrating} with the $n$-qubit Pauli group $C_1$--- are the standard and most convenient non-Clifford elements to enable universal quantum computing when Clifford gates (elements in $C_2$) are available~\cite{gottesman1999demonstrating}. Although any circuit can have an equivalent teleportation gadget, $C_3$ gates can be implemented with the corresponding resource states and conditional operators in the Clifford group, so that the state and gate costs coincide~\cite{gottesman1999demonstrating,zhou2000methodology}.
If the unitaries are additionally diagonal then an explicit teleportation gadget can be given that teleports the gate $\hat{U}$ with the resource state $\hat{U}(\ket{+})^{\otimes n}\equiv\ket{U}$, with $\ket{+}=(\ket{0}+\ket{1})/\sqrt{2}$~\cite{howard2017application}.

This property of the third level of the Clifford hierarchy allows us to bound the $\ket{U}$-cost (or $U$-count) of a target unitary $\hat{U}_{\rm{target}}$ if both gates $\hat{U}$ and $\hat{U}_{\rm{target}}$ belong to $C_3$,
\begin{align}
\label{eq:bound-non_clifford}
   \umag(\ket{U}^{\otimes m }) \leq \umag(\ket{U_{\rm{target}}}) \leq  \umag(\ket{U}^{\otimes m+1 }).
\end{align}
In particular, we estimate the number of $T$ gates (or number of $\ket{H}$ states) needed to implement different unitaries from $C_3$. 
 We quantify the $T$-count, i.e. the number of $T$ gates needed, since the $\{\textrm{Clifford},T\}$ constitutes a universal gate set~\cite{bravyi2005universal}. Equivalently, we can measure the $\ket{H}$-cost ---the number of required $\ket{H} = (\ket{0}+ e^{i\pi/4}\ket{1})/\sqrt{2}$ states--- since the $\ket{H}$ magic state can be consumed to implement the non-Clifford gate $T$~\cite{bravyi2005universal}. 
We analyze the gates characterized with the Robustness of Magic~\cite{howard2017application}, which allowed for improved gate synthesis and proved the optimality of several circuits. The lower bound obtained with the GKP Magic coincides with the lower bound given by the Robustness of Magic for these cases~{\cite[Sec. V]{supmat}}. 

Moreover, we study the multiply-controlled phase gates $\hat{M}_\phi$, which are from the third level of the Clifford hierarchy~\cite{seddon2019quantifying} and have a diagonal representation in computational basis
\begin{align}
\label{eq:M_t}
    {M}_\phi =\text{diag}(1,\dots,1,e^{i \phi}).
\end{align}
They include the multiply-controlled gates $C^{n-1}Z$ ($\phi=\pi$), $C^{n-1}S$ ($\phi=\frac{\pi}{2}$), and $C^{n-1}T$ ($\phi=\frac{\pi}{4}$), where we use the notation for an $n-1$ times controlled $G$ gate as $C^{n-1}G$.

\begin{figure}
    \centering
    \includegraphics{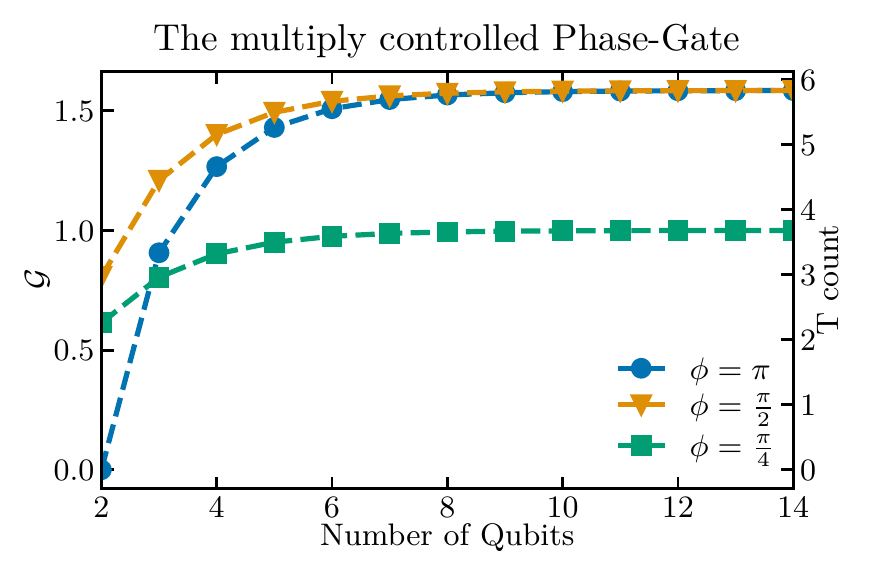}
    \caption{The GKP Magic $\umag(\ket{{M}_{\phi}})$, with $\ket{M_\phi}= \hat{M}_\phi \ket{+}^{\otimes n}$, where $n$ is the number of qubits. The family of unitaries $\hat{M}_\phi$ of Eq.~(\ref{eq:M_t}) belong to diagonal gates of the third level of the Clifford hierarchy, and the state $\ket{M_\phi}$ can be teleported to generate the corresponding gate without additional cost. Here, $\phi=\pi$ corresponds to the $C^{n-1}Z$ gate, $\phi= \frac{\pi}{2}$ to a $C^{n-1}S$ gate and $\phi=\frac{\pi}{4}$ to a $C^{n-1}T$ gate. The GKP Magic converge to a finite value for increasing numbers of qubits.}
    \label{fig:phase}
\end{figure}

We derive analytically the GKP Magic value for any ${M}_\phi$ gate dimension~{\cite[Sec. V]{supmat}}. Fig.~\ref{fig:phase} shows that the GKP Magic for different $\hat{M}_\phi$ gates converges to a finite value as the number of qubits increases. We analyze analytically this asymptotic behavior~{\cite[Sec. V]{supmat}}.
Furthermore, we give analytical expressions for the GKP Magic for the state $\ket{H}$, the quantum adder and the quantum Fourier transform~{\cite[Sec. V]{supmat}}.

Since there exist Clifford operations $\hat{U}$ and $\hat{V}$ such that $C^{n-1}X=\hat{U} C^{n-1}Z \hat{V}$, multiply controlled $X$- and $Z$-gates are expected to contain the same amount of magic. We confirm this numerically~{\cite[Sec. V]{supmat}}, and in particular we compare our obtained $T$-count for the $C^3 Z$ gate with known values to implement a Toffoli-gate $C^3 X$.
The $T$-count provided by our measure coincides with the count of the optimal teleportation gadget of the Toffoli gate~\cite{jones2013low}, but it does not prove the optimality of the $C^{n-1}X$ gate gadget with $4(n-2)$ $T$ gates.
Notice that the optimal unitary circuit to synthesize the Toffoli and Fredkin gates includes $7$ $T$ gates~\cite{mosca2021a-polynomial}, which does not contradict the optimal bound for general synthesis, where measurements and classical feed-forward are allowed. 

In general, we can quantify the GKP Magic of any unitary by using the Choi–Jamiołkowski isomorphism~\cite{choi1975completely, jamiokowski1972linear}, even those that are not diagonal unitaries from the third level of the Clifford hierarchy. With this correspondence, we map unitaries to quantum states suitable for our measure defined in Eq.~(\ref{eq:MM})~{\cite[Sec. V]{supmat}}. The GKP Magic of general unitaries can be used as well to lower bound the resources needed for their implementation. For gates outside $C_3$ we do not expect tight bounds, as one may need to use non-Clifford gates to correct measurements in teleportation schemes or dispose of output states in probabilistic protocols.

Using the Choi–Jamiołkowski isomorphism we calculate numerical values for gates analyzed in~\cite{mosca2021a-polynomial}. The largest system size we consider involves $12$ qubits and corresponds to the $C^{5}X$ gate. The calculated values as well as the most magical two-qubit unitaries and the most magical $3$-qubit states can be found in the Supplemental material~{\cite[Sec. V]{supmat}}.

\paragraph*{GKP Magic and other magic quantifiers.---}

It is interesting to compare the GKP Magic to other magic quantifiers introduced previously in the literature. 
The sum negativity of a discrete Wigner function is a magic monotone for general odd qudit systems~\cite{veitch2014the-resource}. However, the extension to qubit systems remains challenging~\cite{delfosse2015wigner,raussendorf2017contextuality,kocia2017discrete,raussendorf2020phase-space-simulation}.
Even though the discrete Wigner function resembles the Wigner representation of a GKP-encoded qubit restricted to a unit cell, the corresponding magic monotones ---the sum negativity and the GKP Magic--- differ for the single-qubit case~{\cite[Sec. IV]{supmat}}, where both are valid measures.

In contrast to the stabilizer nullity and similarly to the Robustness of Magic, our measure assigns a small value close to stabilizer states. For instance, the GKP Magic of the pure single-qubit state $\ket{\psi}=(\ket{0}+e^{i \phi}\ket{1})/\sqrt{2}$ tends to zero for small angles $\phi \rightarrow 0$. This feature explains the GKP Magic of the multiply-controlled phase gates converging to a finite value, opposite to the stabilizer nullity for the $C^nZ$ gate~\cite{beverland2020lower}. 
Through the GKP magic connection with the stabiliser R\'enyi entropy~\cite{leone2022stabilizer} and st-norm~\cite{howard2017application}, we conclude that it lower bounds the Robustness~\cite{howard2017application} and the stabilizer nullity~\cite[Sec. III]{supmat}.  

Crucially, in contrast to most other magic measures, the computation of our magic measure does not include optimization, while it allows us to find analytical expressions for $n$-qubit states. Also note that the computation of GKP Magic in Eq.~(\ref{eq:MM}) does not require any explicit CV phase space calculations.

Numerically, our measure requires three sums that scale with $2^{3n}$ additions of matrix elements. 
While e.g. the Robustness of Magic could be computed for systems sizes of up to 5 qubits~\cite{howard2017application}, or for product states with specific symmetries~\cite{heinrich2019robustness}, in this work we calculate the GKP Magic for general states of up to $n=12$ qubits. The computation for 12 qubits takes 962s on one core on a laptop CPU (Intel Core i7).
 We expect that larger system sizes are reachable, with 12 qubits not being a hard limit.
 These running times and the additivity of our measure open the possibility of exploring previously unreachable system sizes.

\paragraph*{Discussion and perspective views.---}

In summary, we have introduced a new additive magic measure for multi-qubit pure states, the GKP Magic, derived using bosonic codes ---the GKP encoding--- and considering the WLN in CV systems. Moreover, we have established a connection with the st-norm and stabiliser R\'enyi entropy, with the former initially introduced solely as a one-way magic witness. Crucially, the CV framework allows us to prove the properties of our measure by transferring properties of the WLN. 
The convenient expression of the GKP Magic in Eq.~(\ref{eq:MM}) allows us to lower bound the resources needed for general unitary synthesis and state conversion. In contrast to existing monotones, computing our measure does not require numerical optimization and we can outperform previous results involving up to $\approx 5$ qubits for general states, easily reaching 12-qubit states. Therefore, the GKP magic can be used to address general gate synthesis ---where unitary operations, measurements on auxiliary systems and classical feed-forward are allowed--- and lower bound unitaries and states that were out of reach previously. We also confirm existing optimal lower bounds for several unitary gates studied previously, including the Toffoli gate. Moreover, we have derived analytical expressions of our measure for multiply-controlled phase gates, the quantum adder, and the quantum Fourier transform for an arbitrary number of qubits. Similar to the st-norm and the Robustness of Magic case, we find lower bounds for any multi-qubit state and the general scenario of probabilistic stabilizer protocols using the Wigner negativity per cell, well-defined for mixed states.

 Since the GKP encoding can be applied to qudits of any dimension, it is natural to ask whether we can define a generalized GKP Magic and how it would be related to the discrete Wigner function for odd prime dimensions.
Another interesting open question is weather the GKP magic quantifies the hardness of classical simulation of Clifford computation with additional resource states. 
 Our work sheds new light on magic measures by investigating bosonic codes and CV state conversion. As such, it opens the question as to whether other properties of finite-dimensional systems could be assessed by mapping them to infinite-dimensional ones, and thereby bridging and transferring results from two independent areas of quantum information. Finally, the core idea of connecting concepts of CV and DV systems via bosonic codes can be interpreted as an operational blueprint for resource theories of finite-dimensional systems, beyond quantum computation.

 \let\oldaddcontentsline\addcontentsline
\renewcommand{\addcontentsline}[3]{}

\begin{acknowledgments}
We thank Mark Howard for valuable insights and Timo Hillmann and Robert Jonsson for fruitful discussions.
 G. F. acknowledges support from the Swedish Research Council (Vetenskapsrådet) through the project grant QuACVA. G. F., O. H. and L. G.-\'{A}. acknowledge support from the  Knut and Alice Wallenberg Foundation through the Wallenberg  Center for Quantum Technology (WACQT). L.H. acknowledges financial support from the Gender Initiative for Excellence at Chalmers (GENIE).
 \end{acknowledgments}
 
 \bibliography{bib}

\clearpage
\onecolumngrid
\let\addcontentsline\oldaddcontentsline
\appendix

 \renewcommand{\tocname}{Supplemental Material}

\tableofcontents

\makeatletter
\let\toc@pre\relax
\let\toc@post\relax
\makeatother 

\section{Wigner function analysis of Gottesman-Kitaev-Preskill encoded qubits}

In this section, we derive the Wigner logarithmic negativity per GKP lattice cell. This quantity is the main ingredient to define our magic measure, the GKP magic. First, we give a general expression for the Wigner function of an $n$-qubit state encoded in the GKP code. We then rewrite the Wigner function in a way that makes it possible to read off the value at every point in phase space. The rewritten Wigner function enables us to perform the integral needed to calculate Wigner logarithmic negativity by splitting the integral. Consequently, we obtain a sum of coefficients that we then rewrite in a new compact form.
\subsubsection{Wigner function of $n$-qubit encoded states}
\label{ap:Wigner}
Encoding the information of finite-dimensional quantum systems into bosonic systems is an alternative strategy for quantum error correction. Here, we use the GKP encoding to describe the qubit states with the language of continuous quantum variables~\cite{gottesman2001encoding}. Thus, we encode $n$ qubits in the $2n$-dimensional phase space of $n$ oscillators, by exploiting the translational symmetry of the GKP code. The tensor product of Pauli operators can then be defined as displacements
\begin{align}
\label{eq:CVPauli}
    \hat{U}_{\alpha \beta} = \exp\qty[i\sqrt{2\pi}\qty(\sum_{i=1}^n \alpha_i \hat{p}_i+ \beta_i \hat{q}_i)],
\end{align}
where $\alpha_i, \beta_i$ are real numbers and $\hat{p}_i, \hat{q}_i$ canonical variables that satisfy the relation $\left[\hat{q}_i,\hat{p}_j\right]=i \delta_{ij}$, with $\hbar=1$.
The GKP code space is the simultaneous $+1$ eigenspace of the stabilizer generators, Pauli operators that commute with each other. The Pauli operators of Eq.~(\ref{eq:CVPauli}) commute if $\omega(\bm{\alpha \beta},\bm{\alpha' \beta'}) = \bm{\alpha} \cdot \bm{\beta'} - \bm{\alpha'} \cdot \bm{\beta}$ is an integer, and have a one-to-one correspondence to the points of a lattice in phase space through
\begin{align}
    U(k_1,\dots,k_{2n})= \exp\qty[i\sqrt{2\pi}\qty(\sum_{i=1}^{2n} k_i \bm{a}_i)],
\end{align}
where $\bm{a}_i$ are the $2N$ basis vectors of the lattice, each consisting of linear combinations of $\qn$ and $\pn$, the $n$-tuples $(q_1, \dots,q_n)$ and $(p_1, \dots,p_n)$, and $k_i$ are integers to move from one cell to another. For a more in-depth treatment consult~\cite{gottesman2001encoding}.
 For simplicity, we choose the code space lattice to be a hypercube of side-length $2 \sqrt{\pi}$ generated by stabilizer operators that are linear combinations of $\hat{q}_i$ variables, and separately, combinations of $\hat{p}_i$ variables --- codes of the CSS type~\cite{steane1996error, calderbank1996good}.

\begin{figure*}
    \centering
    \includegraphics{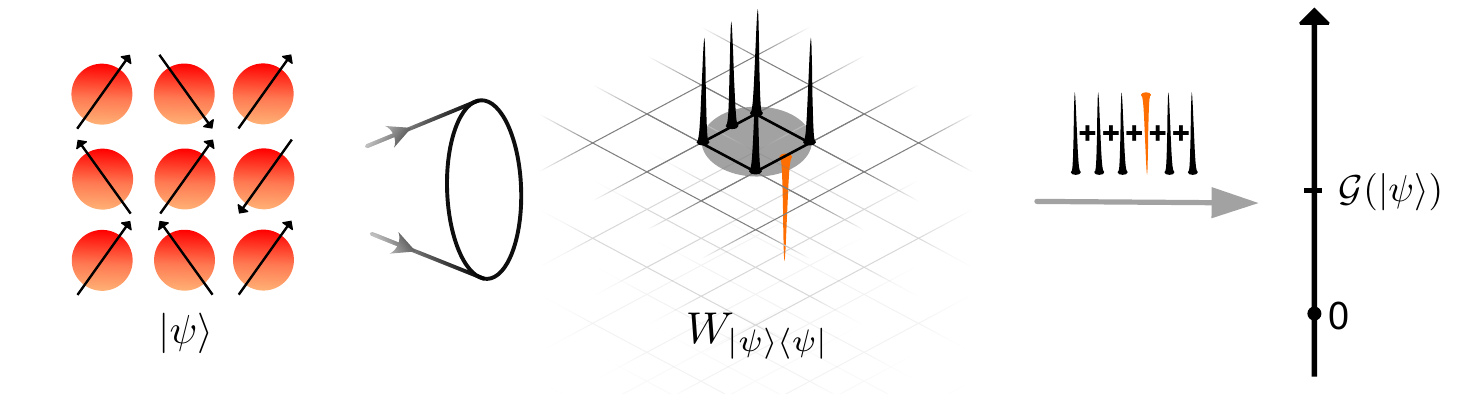}
    \caption{A schematic representation of the GKP magic. First we map a general pure m-qubit state $\ket{\psi}$ into the GKP-encoding, which are $\delta$ peaks on a lattice in the Wigner representation $W_{\ket{\psi}\bra{\psi}}$. By using the translational invariance of the lattice, we only consider one unit cell. We then calculate the Wigner logarithmic negativity of the GKP-encoded states, which is equivalent to sum the weights of the $\delta$ peaks. By doing that we define a magic measure $\umag(\ket{\psi})$ that classifies the distance to stabilizer states.}
    \label{fig:schematics}
\end{figure*}

The idea of our magic measure is to use the GKP encoding and compute the Wigner logarithmic negativity of one cell.
To calculate the Wigner logarithmic negativity of a GKP state, we first need to derive a general expression for the Wigner function.
In this section, we derive the $n$-qubit Wigner function in GKP encoding. The main ideas behind the definition of the GKP magic can be found in Fig.~\ref{fig:schematics}.

A common representation for quantum states and processes is the Wigner functions in a quantum phase space.
The Wigner function is defined as
\begin{align}
    W_{\hat{\sigma}} (\qn,\pn) &= \qty(\frac{1}{2\pi})^n \int_{-\infty}^\infty \dd^n\bm{x} e^{i\pn \bm{x}} \bra{\qn +\frac{\bm{x}}{2}} \hat{\sigma}\ket{\qn-\frac{\bm{x}}{2}}_{\hat{q}},
\end{align}
where $\hat{\sigma}$ is the state of the continuous variable system. In the following we will denote as $\hat{\rho}= \sum_{\bm{u},\bm{v} \in \FF_2^n} \rho_{\bm{u},\bm{v}}\ket{\bm{u}}\bra{\bm{v}}$ the state of the qubit system, to avoid confusion between the density operator of a qubit system and the representation of the qubit system in GKP-encoding.
Starting from this general definition of the $n$-mode Wigner function 
and using the representation of a general $n$-qubit state in GKP encoding, we arrive at 
\begin{align}
    &W_{\hat{\rho}}(\qn,\pn) = \sum_{u,v \in \FF_2^n} \rho_{uv} \prod_{i=1}^n   \frac{1}{2\pi}\int_{-\infty}^\infty \dd{x_i} e^{i p_i x_i}\qty[\sum_{s_i = - \infty}^\infty \delta\qty(q_i+\frac{x_i}{2} - \sqrt{\pi} (u_i + 2s_i) )]    \qty[\sum_{t_i = - \infty}^\infty \delta\qty(q_i-\frac{x_i}{2} - \sqrt{\pi} (v_i + 2t_i) )] ,
\end{align}
where each integral goes only over a single oscillator.
The solution to each of these integrals is calculated in~\cite{garcia-alvarez2021from}.
The $n$-qubit Wigner function in GKP encoding is then given by
\begin{align}
\label{eq:ap_n_qubit_w}
  W_{\hat{\rho}}(\qn,\pn) = \frac{1}{(4\sqrt{\pi})^n}  \sum_{\bm{u},\bm{v} \in \FF_2^n} \rho_{\bm{u},\bm{v}} \prod_{i=1}^n \qty[ \sum_{s_i,t_i}(-1)^{\frac{s_i}{2}(u_i-v_i-2t_i)}    \delta\qty(p_i-\frac{\sqrt{\pi}}{2}  s_i) \delta\qty(q_i-\frac{\sqrt{\pi}}{2}(2t_i+u_i+v_i)) ].
  \end{align}

The Wigner function consists of a sum of Dirac-$\delta$ distribution positioned at all lattice sites in phase space.

\subsubsection{Recursive form of the $n$-qubit encoded states Wigner function}
\label{ap:omega}
To calculate the Wigner logarithmic negativity of a general GKP state, we need to compute integrals over absolute values of sums.
We want to reorder the Wigner function, given in Eq.~(\ref{eq:ap_n_qubit_w}), so that it is possible to extract immediately the coefficients at each lattice site. This change significantly simplifies the integrals needed to calculate the Wigner logarithmic negativity, because we can then split the integration domain. The $n$-qubit Wigner function within one unit cell $q_i,p_i \in [0,2\sqrt{\pi)}$ is given by
\begin{align}
\begin{split}
     (4\sqrt{\pi})^n W_{\hat{\rho}} (\qn,\pn)&=  \sum_{\bm{u},\bm{v} \in \FF_2^n} \rho_{\bm{u},\bm{v}} \prod_{i=1}^n \qty[ \sum_{s_i,t_i}(-1)^{\frac{s_i}{2}(u_i-v_i-2t_i)}    \delta\qty(p_i-\frac{\sqrt{\pi}}{2}  s_i) \delta\qty(q_i-\frac{\sqrt{\pi}}{2}(2t_i+u_i+v_i)) ]\\
      &= \sum_{u_1,v_1 \in \{0,1 \}}\cdots  \sum_{u_n,v_n \in \{0,1 \}} \rho_{u_1,\dots,u_n,v_1,\dots,v_n} \prod_{i=1}^n \qty[\cdots]\\
      &= \sum_{u_1,v_1 \in \{0,1 \}}\cdots  \sum_{u_{n-1},v_{n_1} \in \{0,1 \}} \Bigg[ \sum_{u_n,v_n \in \{0,1 \}} \rho_{u_1,\dots,u_n,v_1,\dots,v_n} \sum_{s_n, t_n} (-1)^{\frac{s_n}{2}(u_n-v_n-2t_n)}  \\
      &\times \delta\qty(p_n-\frac{\sqrt{\pi}}{2}  s_n) \delta\qty(q_n-\frac{\sqrt{\pi}}{2}(2t_n+u_n+v_n))  \Bigg]  \prod_{i=1}^{n-1} \qty[\cdots]\\
      &= \sum_{u_1,v_1\in \{0,1 \} }\cdots  \sum_{u_{n-1},v_{n-1}\in \{0,1 \}}  \Bigg[\sum_{l_n,m_n=0}^3 w_{l_n,m_n}(u_1,\dots,u_{n-1},v_1,\dots,v_{n-1})  \delta\qty(p_n-m_n \frac{\sqrt{\pi}}{2} )  \delta\qty(q_n- l_n \frac{\sqrt{\pi}}{2}) \Bigg]\\
      & \times\; \prod_{i=1}^{n-1} [\cdots]\\
      &=  \sum_{u_1,v_1\in \{0,1 \} }\cdots  \sum_{u_{n-2},v_{n-2}\in \{0,1 \}} \sum_{l_n,m_n=0}^3\Bigg[ \sum_{u_{n-1},v_{n-1}\in \{0,1 \}} w_{l_n,m_n}(u_1,\dots,u_{n-1},v_1,\dots,v_{n-1})  \\   
      &\times  \sum_{s_{n-1}, t_{n-1}} (-1)^{\frac{s_{n-1}}{2}(u_{n-1}-v_{n-1}-2t_{n-1})}\delta\qty(p_{n-1}-\frac{\sqrt{\pi}}{2}  s_{n-1}) \delta\qty(q_{n-1}-\frac{\sqrt{\pi}}{2}(2t_{n-1}+u_{n-1}+v_{n-1})) \Bigg]\\
      &\delta\qty(p_n-m_n \frac{\sqrt{\pi}}{2} )  \delta\qty(q_n- l_n \frac{\sqrt{\pi}}{2})   \prod_{i=1}^{n-2}[\cdots]\\
      &= \sum_{u_1,v_1\in \{0,1 \} }\cdots  \sum_{u_{n-2},v_{n-2}\in \{0,1 \}}\sum_{l_{n-1},m_{n-1}=0}^3  \sum_{l_n,m_n=0}^3  w_{l_{n-1},m_{n-1},l_n,m_n}(u_1,\dots,u_{n-2},v_1,\dots,v_{n-2})  \\
      &\times \delta\qty(p_{n-1}-m_{n-1} \frac{\sqrt{\pi}}{2} )  \delta\qty(q_{n-1}- l_{n-1} \frac{\sqrt{\pi}}{2}) \delta\qty(p_n-m_n \frac{\sqrt{\pi}}{2} )  \delta\qty(q_n- l_n \frac{\sqrt{\pi}}{2})   \prod_{i=1}^{n-2}[\cdots]
      \end{split}
\end{align}
By repeating the last step $n-2$ times, the expression simplifies to $2n$ equally spaced $\delta-$distributions
\begin{align}
\label{eq:equallyspacedW}
     (4\sqrt{\pi})^n W_{\hat{\rho}} (\qn,\pn)= \sum_{l_{1},m_{1}} \cdots \sum_{l_n,m_n}   w_{l_1,m_1,\dots,l_n,m_n} \delta\qty(p_{1}-m_{1} \frac{\sqrt{\pi}}{2} )  \delta\qty(q_{1}- l_{1} \frac{\sqrt{\pi}}{2})\cdots\;  \delta\qty(p_n-m_n \frac{\sqrt{\pi}}{2} )  \delta\qty(q_n- l_n \frac{\sqrt{\pi}}{2}),
\end{align}
where we define $w$ of first order as
\begin{align}
\label{eq:w_first}
      w_{0,m_n} (u_1,\dots,u_{n-1},v_1,\dots,v_{n-1})&=  \rho_{u_1,\dots,0,v_1,\dots,0}  + (-1)^{m_n} \rho_{u_1,\dots,1,v_1,\dots,1} \nonumber \\
      w_{1,m_n} (u_1,\dots,u_{n-1},v_1,\dots,v_{n-1})&=  (-i)^{m_n} \rho_{u_1,\dots,0,v_1,\dots,1} + i^{m_n} \rho_{u_1,\dots,1,v_1,\dots,0} \nonumber \\
      w_{2,m_n} (u_1,\dots,u_{n-1},v_1,\dots,v_{n-1})&=   (-1)^{m_n} \rho_{u_1,\dots,0,v_1,\dots,0}+  \rho_{u_1,\dots,1,v_1,\dots,1} \nonumber \\
      w_{3,m_n}(u_1,\dots,u_{n-1},v_1,\dots,v_{n-1}) &= i^{m_n} \rho_{u_1,\dots,0,v_1,\dots,1} + (-i)^{m_n} \rho_{u_1,\dots,1,v_1,\dots,0}
\end{align}
and of $j$-th order recursively as 
\begin{align}
\label{eq:w_ith}
      w_{0,m_j,\dots,l_n,m_n} (u_1,\dots,u_{j-1},v_1,\dots,v_{j-1})=& \; w_{l_{j+1},m_{j+1},\dots,l_n,m_n} (u_1,\dots,u_{j-1},0,v_1,\dots,v_{j-1},0) \nonumber \\
      &+ (-1)^{m_j} w_{l_{j+1},m_{j+1},\dots,l_n,m_n} (u_1,\dots,u_{j-1},1,v_1,\dots,v_{j-1},1) \nonumber\\
     w_{1,m_j,\dots,l_n,m_n} (u_1,\dots,u_{j-1},v_1,\dots,v_{j-1})=& \; (-i)^{m_j} w_{l_{j+1},m_{j+1},\dots,l_n,m_n} (u_1,\dots,u_{j-1},0,v_1,\dots,v_{j-1},1) \nonumber \\
     &+ i^{m_j} w_{l_{j+1},m_{j+1},\dots,l_n,m_n} (u_1,\dots,u_{j-1},1,v_1,\dots,v_{j-1},0) \nonumber \\
     w_{2,m_j,\dots,l_n,m_n} (u_1,\dots,u_{j-1},v_1,\dots,v_{j-1})=& \;  (-1)^{m_j} w_{l_{j+1},m_{j+1},\dots,l_n,m_n} (u_1,\dots,u_{j-1},0,v_1,\dots,v_{j-1},0) \nonumber \\
     &+  w_{l_{j+1},m_{j+1},\dots,l_n,m_n} (u_1,\dots,u_{j-1},1,v_1,\dots,v_{j-1},1) \nonumber \\
     w_{3,m_j,\dots,l_n,m_n}(u_1,\dots,u_{j-1},v_1,\dots,v_{j-1}) =& \; i^{m_j} w_{l_{j+1},m_{j+1},\dots,l_n,m_n} (u_1,\dots,u_{j-1},0,v_1,\dots,v_{j-1},1) \nonumber \\
     &+ (-i)^{m_j} w_{l_{j+1},m_{j+1},\dots,l_n,m_n} (u_1,\dots,u_{j-1},1,v_1,\dots,v_{j-1},0).
\end{align}
All the parameters $w_{l_1,m_1,\dots,l_n,m_n}$ of Eq.~(\ref{eq:equallyspacedW}) can therefore be calculated by this recursion. This can be generalized easily to the entire phase space by using the translational invariance of the Wigner function.
Note that this is a generalization of work done for one qubit in~\cite{garcia-alvarez2021from}.
\subsubsection{Integrals in the Wigner logarithmic negativity of $n$-qubit encoded states}
\label{ap:Integralsplit}
We are interested in calculating the Wigner logarithmic negativity of the $n$-qubit Wigner function given by Eq.~(\ref{eq:ap_n_qubit_w}) to define a new magic measure.
In this section, we go over how we carried out the integrations.
Following the definition of the Wigner logarithmic negativity ~\cite{albarelli2018resource, takagi2018convex} 
\begin{align}
\label{eq:ap_Wigner_log_neg}
    \logneg(\hat{\sigma}) = \log_2 \qty( \int_{-\infty}^\infty \ddqn \ddpn  |W_{\hat{\sigma}}(\qn,\pn)|),
\end{align}
we have to integrate over integrals of the form defined in 
\begin{align}
    \int \dd{q} |f(q) \delta(q) |.
\end{align}
The Dirac-$\delta$ distribution can be represented as the threshold of a function sequence. Let us choose a representation that is a Dirac sequence and thus is positive everywhere, as is the case for the representation as a Gaussian
\begin{align}
\lim_{\epsilon \rightarrow 0}\delta_{\epsilon}(x) = \lim_{\epsilon \rightarrow 0} \frac{1}{\sqrt{2\pi \epsilon}} \exp\qty( -\frac{x^2}{2\epsilon}).
\end{align}
Using this representation, the function sequence then gives the Dirac-$\delta$ distribution in the limit
\begin{align}
    \lim_{\epsilon \rightarrow 0}  \int \dd{x} \delta_{\epsilon}(x) f(x) = \int \dd{x} \delta(x) f(x) = f(0).
\end{align}

We define the absolute value of the Dirac-$\delta$ distribution over the action on the representation as a function sequence, namely
\begin{align}
\int \dd{x} \abs{\delta(x) f(x)} =    \lim_{\epsilon \rightarrow 0}  \int \dd{x} \abs{\delta_{\epsilon}(x) f(x)}.
\end{align}
Thus, the threshold should not change if we take absolute values of the Dirac-$\delta$ distribution, since it is positive everywhere.
Therefore, we evaluate the integrals
\begin{align}
    \int \dd{q} |f(q) \delta(q) | = \int \dd{q} |f(q)| | \delta(q) | =  \int \dd{q} |f(q)|  \delta(q) 
\end{align}
For the last equality we considered the delta-distribution in the function sequence representation.
Then it holds for a sum of the form
\begin{align}
\begin{split}
        \int_{-\infty}^\infty \dd{\bm{x}}\abs{ \sum_i \alpha_i  \delta\qty(\bm{x} - \bm{x}_i)} &= \sum_j \int_{\bm{x}_j -\Delta_j}^{\bm{x}_j +\Delta_j}    \dd\bm{x}  \abs{ \sum_i \alpha_i  \delta\qty(\bm{x} - \bm{x}_i)}  \\
        &= \sum_j \int_{\bm{x}_j -\Delta_j}^{\bm{x}_j +\Delta_j}    \dd\bm{x}  \abs{ \alpha_j  \delta\qty(\bm{x} - \bm{x}_j)}  \\
        &= \sum_j \int_{\bm{x}_j -\Delta_j}^{\bm{x}_j +\Delta_j}    \dd\bm{x}  \abs{ \alpha_j}  \delta\qty(\bm{x} - \bm{x}_j)  \\
        &= \sum_j \abs{\alpha_j},
\end{split}
\end{align}
where $\Delta_j$ is a finite value such that the integral domains are not overlapping and only one Dirac-$\delta$ distribution has support and $\alpha_i \in \mathbb{R}$. We will show this for a special case in the following.

Specific for our case, we rewrite the integral by splitting up the integration domain, where only one of the Dirac-$\delta$ distributions has support.
For simplicity we only consider the integration over $q$, integration over both $q$ and $p$ is analogous. We use the notation $-\epsilon$ to symbolize that the value is not in the integral domain, i.e. that $\int_0^{2\sqrt{\pi}-\epsilon} \dd{q} $ goes over the domain $\big[0,2\sqrt{\pi}\big)  $.
\begin{align}
\begin{split}
    &\int_0^{2\sqrt{\pi}-\epsilon} \dd{q}     \qty|\sum_{l=1}^3    \alpha_l\; \delta\qty(q-l\frac{\sqrt{\pi}}{2})| \\
    &=   \int_0^{2\sqrt{\pi}-\epsilon} \dd{q}     \qty|   \alpha_1\; \delta\qty(q-\frac{\sqrt{\pi}}{2}) + \alpha_2\; \delta\qty(q-2\frac{\sqrt{\pi}}{2}) +\alpha_3\; \delta\qty(q-3\frac{\sqrt{\pi}}{2})|\\
    &=\int_0^{\frac{3}{4}\sqrt{\pi}-\epsilon} \dd{q}     \qty|   \alpha_1\; \delta\qty(q-\frac{\sqrt{\pi}}{2}) + \alpha_2\; \delta\qty(q-2\frac{\sqrt{\pi}}{2}) +\alpha_3 \;\delta\qty(q-3\frac{\sqrt{\pi}}{2})|\\
    &+\int_{\frac{3}{4}\sqrt{\pi}}^{\frac{5}{4}\sqrt{\pi}-\epsilon} \dd{q}     \qty|   \alpha_1\; \delta\qty(q-\frac{\sqrt{\pi}}{2}) + \alpha_2\; \delta\qty(q-2\frac{\sqrt{\pi}}{2}) +\alpha_3\; \delta\qty(q-3\frac{\sqrt{\pi}}{2})|\\
    &+\int_{\frac{5}{4}\sqrt{\pi}}^{2\sqrt{\pi}-\epsilon} \dd{q}     \qty|   \alpha_1\; \delta\qty(q-\frac{\sqrt{\pi}}{2}) + \alpha_2\; \delta\qty(q-2\frac{\sqrt{\pi}}{2}) +\alpha_3 \;\delta\qty(q-3\frac{\sqrt{\pi}}{2})|\\
    &=\int_0^{\frac{3}{4}\sqrt{\pi}-\epsilon} \dd{q}\qty|   \alpha_1\; \delta\qty(q-\frac{\sqrt{\pi}}{2}) | + \int_{\frac{3}{4}\sqrt{\pi}}^{\frac{5}{4}\sqrt{\pi}-\epsilon} \dd{q}\qty|   \alpha_2\; \delta\qty(q-\frac{2\sqrt{\pi}}{2}) | +\int_{\frac{5}{4}\sqrt{\pi}}^{2\sqrt{\pi}-\epsilon} \dd{q}\qty|   \alpha_3\; \delta\qty(q-3\frac{\sqrt{\pi}}{2}) |\\
      &= \int_0^{\frac{3}{4}\sqrt{\pi}-\epsilon} \dd{q}\qty|   \alpha_1 |\;\delta\qty(q-\frac{\sqrt{\pi}}{2})  + \int_{\frac{3}{4}\sqrt{\pi}}^{\frac{5}{4}\sqrt{\pi}-\epsilon} \dd{q}\qty|   \alpha_2 |\;\delta\qty(q-\frac{2\sqrt{\pi}}{2})  +\int_{\frac{5}{4}\sqrt{\pi}}^{2\sqrt{\pi}-\epsilon} \dd{q}\qty|   \alpha_3 |\;\delta\qty(q-3\frac{\sqrt{\pi}}{2}) \\
    &= \int_0^{2\sqrt{\pi}-\epsilon} \dd{q}    \sum_{l=1}^3   \qty| \alpha_l| \;\delta\qty(q-l\frac{\sqrt{\pi}}{2}) = \sum_{l=1}^3   \qty| \alpha_l|.
    \end{split}
\end{align}

\subsubsection{Wigner logarithmic negativity of $n$-qubit encoded states}
\label{ap:sum}

We calculate in this section the Wigner logarithmic negativity, using the expression for the general $n$-qubit Wigner function and the technique to evaluate the integrals. Furthermore we will reformulate the recursions defined in Eqs.~(\ref{eq:w_first}) and (\ref{eq:w_ith}) to retrieve a more compact formula for the Wigner negativity of one GKP unit-cell.

We restrict ourselves to one unit cell for a hypercubic lattice $q_i\in\big[0,2\sqrt{\pi}\big) ,p_i\in\big[0,2\sqrt{\pi}\big) $, where we call this domain $\mathcal{C}$. Then the Wigner logarithmic negativity is defined as
\begin{align}
    \logneg_C(\hat{\rho}) = \log_2 \qty( \int_{\mathcal{C}} \ddqn \ddpn \; |W_{\hat{\rho}}(\qn,\pn)|).
\end{align}
As we have shown previously, the Wigner logarithmic negativity of the $n$-qubit GKP-state, defined in Eq.~(\ref{eq:equallyspacedW}), is just a sum of absolute values of the coefficients $w_{l_1,\dots,l_n,m_1,\dots,m_n}$
\begin{align*}
    \logneg_C(\hat{\rho}) = \log_2 \qty( \frac{1}{(4\sqrt{\pi})^n} \sum_{l_1,m_1= 0}^3 \cdots\sum_{l_n,m_n= 0}^3 \abs{w_{l_1,\dots,l_n,m_1,\dots,m_n}} ).
\end{align*}
As was mentioned before, the GKP magic is defined for pure states. We still write the more general density operator $\hat{\rho}$ to indicate that the results in the following sections hold for the more general case.
\begin{table*}
\centering

\begin{tabular}{|c| c c c c|}
\hline
\backslashbox{$l_1$}{$m_1$} & 0&1&2&3\\ \hline 
0 & $ \mathds{1}$  & $ Z$ & $\mathds{1}$ & $Z$\\
1& $X$ & $Y$ & $-X$ & $-Y$\\
2& $\mathds{1}$ & $-Z$ &$\mathds{1}$ & $-Z$\\
3& $X$ & $-Y$ &$-X$ & $Y$\\
\hline
\end{tabular}
\caption{Matrices  $W_{l,m}$ for the single qubit case.}
\label{tab: matrices_1}
\end{table*}
We define the short hand notation
\begin{align*}
    \Vec{A}\cdot \Vec{B} = A\big[B\big].
\end{align*}
where $\Vec{A}$ is the row vectorization of matrix $A$.
Additionally we define the block notation
\begin{align}
    B^n = \begin{pmatrix}
  B^{n-1}_{00} & \rvline &B^{n-1}_{01} \\\hline
B^{n-1}_{10} & \rvline & B^{n-1}_{11}\\
    \end{pmatrix} 
\end{align}
and 
\begin{align}
    B^1_{\bm{i},\bm{j}} = \begin{pmatrix}
  \rho_{\bm{i}0,\bm{j}0} & \rvline &\rho_{\bm{i}0,\bm{j}1} \\\hline
\rho_{\bm{i}1,\bm{j}0} & \rvline & \rho_{\bm{i}1,\bm{j}1}\\
    \end{pmatrix} 
\end{align}
where $\bm{i},\bm{j} \in \FF_2^{n-1}$.
We define the matrices $W_{l,m}$
 \begin{align}
 W_{l,m} = i^{l \cdot m} Z^{m \text{mod}2 } X^{l \text{mod}2 }
 \end{align}
If we look at our definition of $W_{l,m}$ in Table~\ref{tab: matrices_1}, we see that the matrices repeat themselves with different global signs. The global signs are negligible, since we are summing absolute values of our coefficients $w$. Thus we get a multiplicity of $4^n$ if we only consider all the elements of the $n$-qubit Pauli group $C_1$, the first level of the Clifford hierarchy~\cite{gottesman1999demonstrating}.

Using our notation, the leftmost matrix $W_{l,m}$ will act on the entries of block $B^n$ where the next one in then entries of $B^{n-1}$, etc.. The rightmost $W_{l,m}$  will then act on the entries of $B^1$, which are then different elements of the quantum state.

With this notation, we can simplify the Wigner negativity by rewriting the sum of the recursion formulas defined in Sec.~\ref{ap:omega} to
\begin{align}
\begin{split}
  \sum_{l_1,m_1= 0}^1\cdots\sum_{l_n,m_n= 0}^1 \abs{w_{l_1,\dots,l_n,m_1,\dots,m_n}} &= \sum_{l_1,m_1= 0}^1\cdots\sum_{l_{n-1},m_{n-1}= 0}^1 \sum_{l_n, m_n =0}^1 \abs{W_{l_n,m_n}[B^n ]} \\
  &=\sum_{l_1,m_1= 0}^1\cdots\sum_{l_{n-2},m_{n-2}}^1 \sum_{l_n, m_n =0}^1 \sum_{l_{n-1}, m_{n-1} =0}^1 \abs{W_{l_n,m_n}[W_{l_{n-1},m_{n-1}} [B^{n-1}_{i,j}] ]} \\
    &= \sum_{l_1,m_1= 0}^1\cdots\sum_{l_{n},m_{n}}^1 \abs{W_{l_n,m_n}[\dots[ W_{l_1,m_1}   [ B^1_{l,m} ] ]\dots    ]}.\end{split}
\end{align}
We can evaluate the action of the matrices on a block
\begin{align}
\begin{split}
     \abs{\mathds{1}\big[B^n\big]} &=     \abs{\big[B^{n-1}_{00}\big] +  \big[[B^{n-1}_{11}\big]}=\abs{\sum_{i=0}^1  \big[B^{n-1}_{ii}\big]}\\
      \abs{Z \big[B^n\big]}&=  \abs{\big[B^{n-1}_{00}\big] -   \big[B^{n-1}_{11}\big]}=\abs{\sum_{i=0}^1 \qty(-1)^i \big[B^{n-1}_{ii}\big]}\\
     \abs{ X \big[B^n\big]}&=  \abs{\big[B^{n-1}_{01}\big] +   \big[B^{n-1}_{10}\big]}=\abs{\sum_{i=0}^1  \big[B^{n-1}_{ii+1}\big]}\\
      \abs{Y\big[B^n\big]}&=  \abs{\big[B^{n-1}_{01}\big] -   \big[B^{n-1}_{10}\big]}=\abs{\sum_{i=0}^1  \qty(-1)^i \big[B^{n-1}_{ii+1}\big]}.\end{split}
\end{align}
This can be summarized in
\begin{align}
    \sum_{l_n,m_m=0}^1 \abs{ W_{l_n,m_m}   [ B^n ]}= \sum_{i_n,j_n=0}^1 \abs{\sum_{k_n=0}^1   \qty(-1)^{i_n\cdot k_n}  B_{k , k+j}^{n-1} }.
\end{align}

By repeatably applying the matrices on the blocks, we finally arrive at compact expression to calculate the Wigner negativity
\begin{align}
\begin{split}
  \sum_{l_1,m_1= 0}^1\cdots\sum_{l_n,m_n= 0}^1 \abs{w_{l_1,\dots,l_n,m_1,\dots,m_n}} &= \sum_{i_n,j_n=0}^1\cdots\sum_{i_1,j_1=0}^1 \abs{\sum_{k_n=0}^1  \cdots \sum_{k_1=0}^1  \qty(-1)^{i_n\cdot k_n}\dots\qty(-1)^{i_1\cdot k_1}  B_{k_1\dots k_n , k_1+j_1\dots k_n+j_n}^{1} }\\
    &=\sum_{\bm{i},\bm{j} \in \FF_2^n}  \abs{ \sum_{\bm{k} \in \FF_2^n} \qty(-1)^{\bm{i}\cdot \bm{k} }   \rho_{\bm{k},\bm{k}+\bm{j}} },\end{split}
    \label{eq_sum}
\end{align}
where $\rho_{\bm{k}, \bm{k}+\bm{j}}$ are the matrix elements of the $n$-qubit density operator in $\sigma_z$ eigenbasis.
$\bm{i}\cdot \bm{k} = \sum_{j=1}^n  i_j k_j \mod 2$ is the standard binary inner product, and $\bm{k}+\bm{j}$ the bitwise sum $(\bm{k}+\bm{j})_i= k_i + j_i \mod 2 $. 
So if we restrict $\hat{\rho}=\ket{\psi}\bra{\psi}$ with $\ket{\psi} = \sum_{ \bm{i} \in \FF_2^n}  c_{\bm{i}} \ket{\bm{i}}$ as a pure state then the coefficients are given by $\rho_{\bm{k},\bm{k}+\bm{j}} = c^*_{\bm{k}+\bm{j}} c_{\bm{k}} $. Note that we have shown here equivalently a different form of the coefficients in Eq.~(\ref{eq:equallyspacedW}) and the connection to Sec.~\ref{ap:equivalent}.

\section{GKP magic properties}
\label{ap:properties}
A magic measure needs to fulfill a list of properties.
In this section, we prove that the GKP magic is an additive magic measure.

We will start with definitions and  the translation symmetry of the Wigner function of general qubit states encoded in GKP.
Let's consider a general lattice.
Then we can  decompose vectors in lattice vectors $\bm{R}$, that are vectors that translate between unit cells, and vectors within a unit cell $ \bm{r}$
\begin{align}
\begin{split}
    \bm{R} &= \sum_i n_i \bm{a}_i\\
    \bm{r} &= \bm{l} + \bm{R}
    \end{split}
\end{align}
where $\bm{a}_i$ are the basis vectors that can be used to span the entire lattice, $n_i \in \mathcal{Z}$ and $\bm{l}$ a vector within a lattice cell.
A lattice periodic function is defined as
\begin{align}
    f(\bm{r}) = f(\bm{r}+\bm{R}).
\end{align}
Our Wigner function is defined on a hypersquare lattice and thus is lattice periodic as well
\begin{align}
    W_{\hat{\rho}}(\bm{r}) = W_{\hat{\rho}}(\bm{r}+\bm{R}),
\end{align}
where $\bm{r},\bm{R}$ are the vectors containing both quadratures.
Note that for the GKP magic
we consider $\hat{\rho}= \ket{\psi}\bra{\psi}$ a pure state, but use the notation for a density operator to be consistent with the definition used to calculate the Wigner function. The results for the properties of the Wigner negativity hold for mixed qubit states as well.

We call \begin{align}
      \cellneg(\hat{\rho}) =\int_{-\infty}^\infty \ddrn\; \abs{W_{\hat{\rho}} (\bm{r}) } 
\end{align}
the Wigner negativity and
\begin{align}
      \cellneg_{C}(\hat{\rho}) =\int_{\mathcal{C}}\ddrn\; \abs{W_{\hat{\rho}} (\bm{r}) } 
\end{align}
the Wigner negativity of one unit cell.
 
We use the Wigner negativity instead of the Wigner logarithmic negativity $\logneg$ to prove the properties of the GKP magic for brevity. Because of the monotonicity of the logarithm, the proofs apply for the Wigner logarithmic negativity analogously.

Thus for calculating the Wigner negativity, the integrals can be decomposed as

\begin{align}
\begin{split}
    \cellneg(\hat{\rho}) &= \int_{-\infty}^\infty \ddrn\; \abs{W_{\hat{\rho}} (\bm{r}) } \\
     &= \sum_{n_1=-\infty}^\infty \cdots \sum_{n_{2N}=-\infty}^\infty  \int_{\bm{R}_{n_1 n_2}}^{\bm{R}_{n_1 n_2}+ \bm{R}_{n_1=1,n_2=1}} \dd\bm{r}_1 \dots
     \int_{\bm{R}_{n_{2N-1} n_{2N}}}^{\bm{R}_{n_{2N-1} n_{2N}}+ \bm{R}_{n_{2N-1}=1,n_{2N}=1}} \dd\bm{r}_N \abs{W_{\hat{\rho}} (\bm{r} )}\\
     &=\sum_{n_1=-\infty}^\infty \cdots \sum_{n_{2N}=-\infty}^\infty  \int_{0}^{ \bm{R}_{n_1=1,n_2=1}} \dd\bm{r}_1 \dots
     \int_{0}^{\bm{R}_{n_{2N-1}1,n_{2N}=1}} \dd\bm{r}_N \abs{W_{\hat{\rho}} (\bm{r}+\bm{R}_{n_1,\dots,n_{2N}} )}\\
     &=\sum_{n_1=-\infty}^\infty \cdots \sum_{n_{2N}=-\infty}^\infty  \int_{0}^{ \bm{R}_{n_1=1,n_2=1}} \dd\bm{r}_1 \dots
     \int_{0}^{\bm{R}_{n_{2N-1}=1,n_{2N}=1}} \dd\bm{r}_N \abs{W_{\hat{\rho}} (\bm{R} )}\\
     &= \lim_{n_1\rightarrow \infty} \cdots \lim_{n_{2N}\rightarrow \infty} n_1\dots n_{2N} \int_{0}^{ \bm{R}_{n_1=1,n_2=1}} \dd\bm{r}_1 \dots
     \int_{0}^{\bm{R}_{n_{2N-1}=1,n_{2N}=1}} \dd\bm{r}_N \abs{W_{\hat{\rho}} (\bm{r} )}\\
     &= \lim_{n_1\rightarrow \infty} \cdots \lim_{n_{2N}\rightarrow \infty} n_1\cdots n_{2N} \cdot \; \; \cellneg_{C}(\hat{\rho})
     \end{split}
\end{align}
where we use the shorthand notation $\bm{R}_{n_1\dots n_{2N}} =\sum_{i=1}^{2N} n_i \Vec{a}_i $.

In the following derivations, we will use the shorthand notation
\begin{align}
  \lim_{n_1\rightarrow \infty} \cdots \lim_{n_{2N}\rightarrow \infty} n_1\dots n_{2N} \int_{\mathcal{C}}\ddrn\; \abs{W_{\hat{\rho}} (\bm{R}) } = \lim_{n_1\rightarrow \infty} \cdots \lim_{n_{2N}\rightarrow \infty} n_1 \dots n_{2N} \int_{0}^{ \bm{R}_{n_1=1,n_2=1}} \dd\bm{r}_1 \dots
     \int_{0}^{\bm{R}_{n_{2N-1}=1,n_{2N}=1}} \dd\bm{r}_N \abs{W_{\hat{\rho}} (\bm{R} )}.
\end{align}
The results holds as well if we take the logarithm, since $\log(x)$ is a monotonous function for $x$.

\subsubsection{Invariance under Clifford unitaries}

We need to show that the Wigner negativity for one cell is invariant under Clifford unitaries $ \hat{U}_C \subset \hat{U}_G$ on the code space
\begin{align}
    \cellneg_C(\hat{\rho}) = \cellneg_C( \hat{U}_C  \hat{\rho} \hat{U}_C^\dagger).
\end{align}
The Clifford unitaries map a code word to a different code word and leave the lattice thus invariant. It is known that the Wigner negativity is invariant under the action of Gaussian operations~\cite{albarelli2018resource}.
Thus it needs to hold
\begin{align}
\begin{split}
    \int_{-\infty}^\infty \ddrn \abs{W_{\hat{\rho}}(\bm{r})}&= \lim_{n_1\rightarrow \infty} \cdots \lim_{n_{2N}\rightarrow \infty} n_1\dots n_{2N} \int_{\mathcal{C}}\ddrn  \abs{W_{\hat{\rho}}(\bm{r})}\\
    &=\lim_{n_1\rightarrow \infty} \cdots \lim_{n_{2N}\rightarrow \infty} n_1 \dots n_{2N} \int_{\mathcal{C}} \ddrn \abs{W_{( \hat{U}_C  \hat{\rho} \hat{U}_C^\dagger)}(\bm{r})} = \int_{-\infty}^\infty \ddrn \abs{W_{(\hat{U}_C  \hat{\rho} \hat{U}_C^\dagger)}(\bm{r})}
    \end{split}
\end{align}
where we used that the Wigner negativity is invariant under Gaussian unitaries. Thus the value of one cell is invariant under Clifford unitaries as well.

Since our measure is defined as
\begin{align}
\begin{split}
\label{eq:GKPmana-second-expression}
  \umag(\ket{\psi}) &= \log_2 \qty( \int_{\mathcal{C}} \ddqn \ddpn \; |W_{\ket{\psi}\bra{\psi}}(\qn,\pn)|) -n\log_2\qty(\frac{2}{\sqrt{\pi}})\\
    &= \log_2 \qty(  \cellneg_C(\ket{\psi}\bra{\psi})) -n\log_2\qty(\frac{2}{\sqrt{\pi}}),
  \end{split}
\end{align}
the GKP magic is invariant under Clifford unitaries as well. The offset $\cellnegint(n)=\qty(\frac{2}{\sqrt{\pi}})^n$ is calculated in Sec.~\ref{ap:faith}.

\subsubsection{Additivity}
We show that the GKP magic is additive since the Wigner negativity of one cell is multiplicative. It holds that
\begin{align}
\begin{split}
    \cellneg(\hat{\rho}_1 \otimes \hat{\rho}_2) &= \lim_{n_1\rightarrow \infty} \cdots \lim_{n_{2N}\rightarrow \infty} n_1\dots n_{2N}
    \lim_{m_1\rightarrow \infty} \cdots \lim_{m_{2N}\rightarrow \infty} m_1\dots m_{2N}\int _{\mathcal{C}_1 \times \mathcal{C}_2} \ddrn \abs{W_{\hat{\rho}_1\otimes \hat{\rho}_2}(\bm{r})}\\
        &= \lim_{n_1\rightarrow \infty} \cdots \lim_{n_{2N}\rightarrow \infty} n_1\dots n_{2N}
    \lim_{m_1\rightarrow \infty} \cdots \lim_{m_{2N}\rightarrow \infty} m_1\dots m_{2N} \int_{\mathcal{C}_1}\ddrn_1 \abs{W_{\hat{\rho}_1}(\bm{r}_1)}\int_{\mathcal{C}_2 }\ddrn_2 \abs{W_{\hat{\rho}_2}(\bm{r}_2)}\\
        &=  \cellneg(\hat{\rho}_1)  \cellneg(\hat{\rho}_2)\end{split}
\end{align}
where we used that the Wigner function factorizes for product states. Thus the  Wigner negativity of one cell is multiplicative and the GKP magic additive, because the offset can be split accordingly as well
\begin{align}
  \umag(\ket{\psi_1}\otimes \ket{\psi_2}) &= \log_2 \qty( \int_{\mathcal{C}} \ddqn \ddpn \; \abs{W_{\ket{\psi_1,\psi_2}\bra{\psi_1,\psi_2}}(\qn,\pn)}) -(n_1+n_2)\log_2\qty(\frac{2}{\sqrt{\pi}})\\
    &= \log_2 \qty(  \cellneg_C(\ket{\psi_1,\psi_2}\bra{\psi_1,\psi_2})) -(n_1+n_2)\log_2\qty(\frac{2}{\sqrt{\pi}})\\
    &= \log_2 \qty(  \cellneg_C(\ket{\psi_1}\bra{\psi_1})) -n_1\log_2\qty(\frac{2}{\sqrt{\pi}}) +  \log_2 \qty(  \cellneg_C(\ket{\psi_2}\bra{\psi_2})) -n_2\log_2\qty(\frac{2}{\sqrt{\pi}})\\
    &= \umag(\ket{\psi_1}) + \umag( \ket{\psi_2}).
\end{align}

\subsubsection{Faithfulness}
\label{ap:faith}
Here, we prove that the GKP magic is a faithful monotone. That is, we want to enforce that $\umag(\ket{\psi_S})=0$ if and only if $\ket{\psi_S}$ is a stabilizer state. The GKP magic of an $n$-qubit pure state is defined in terms of the Wigner negativity per cell $\cellneg_C$ of the corresponding GKP encoded state, as shown in Eq.~(\ref{eq:GKPmana-second-expression}). From the definition, we observe that the measure is faithful if we find a normalization $\cellnegint(n)$ for $n$-qubit systems such that
\begin{align}
    \frac{1}{\cellnegint(n)}\cellneg_C(\ket{\psi_S}\bra{\psi_S}) =1
\end{align}
if and only if $\ket{\psi_S}$ is a stabilizer state.

Firstly, we consider single-qubit systems, $n=1$. As shown in Sec.~\ref{ap:polytope}, the Wigner negativity per cell of single-qubit stabilizer states $\ket{\psi_S}$ is given by~\cite{garcia-alvarez2021from}
\begin{align}
    \cellneg_C(\ket{\psi_S}\bra{\psi_S}) = \frac{2}{\sqrt{\pi}}.
\end{align}
Therefore, the normalization constant that guarantees that the GKP magic is zero for stabilizer states is 
\begin{align}
    \cellnegint(1)=\frac{2}{\sqrt{\pi}}.
\end{align}
We also need to show that the GKP magic vanishes only for stabilizer states. That is, that the Wigner negativity per cell is $2/\sqrt{\pi}$ only if the single-qubit state is a stabilizer state. We consider a general single-qubit state $\ket{\psi} = \cos(\frac{\theta}{2})\ket{0} + \sin(\frac{\theta}{2}) e^{i\phi} \ket{1}$ and its Wigner negativity per cell~\cite{garcia-alvarez2021from}
\begin{equation}
    \cellneg_C(\ket{\psi}\bra{\psi}) = \frac{2}{\sqrt{\pi}}\qty(\abs{\cos(\theta)}+\abs{\sin(\theta) \cos(\phi)} +\abs{\sin(\theta) \sin(\phi)}),
\end{equation}
which becomes $2/\sqrt{\pi}$ for qubit states characterized by $\theta$ and $\phi$ values satisfying
\begin{align}    
    \frac{2}{\sqrt{\pi}} = \frac{2}{\sqrt{\pi}}\qty(\abs{\cos(\theta)}+\abs{\sin(\theta) \cos(\phi)} +\abs{\sin(\theta) \sin(\phi)}).
\end{align}
The solutions are given by $\theta = k \pi$ for any integer $k$ and arbitrary $\phi$; and by $\theta = (2k+1)\frac{\pi}{2}$ and $\phi=0$ or $\phi=(2\ell+1) \frac{\pi}{2}$ for any integers $k$ and $\ell$. These parameter values represent single-qubit stabilizer states, which proves the GKP magic faithfulness for the single-qubit pure states.

Now, we address the general $n$-qubit systems case. First, we prove that if a state $\ket{\psi_S}$ is a stabilizer state, then $\umag(\ket{\psi_S})=0$. Multi-qubit stabilizer states are constructed via tensor product of single-qubit stabilizer states and the application of Clifford unitary operations. Since Clifford operations do not increase the Wigner negativity per cell $\cellneg_C$, we can consider the normalization $\cellnegint(n)$ for $n$-qubit product states without loss of generality. Taking into account the multiplicativity of the Wigner negativity, the normalization for $n$-qubit systems is given by
\begin{align}
    \cellnegint(n)= \qty(\frac{2}{\sqrt{\pi}})^n.
\end{align}
This normalization is included in the definition of the GKP magic in Eq.~(\ref{eq:GKPmana-second-expression}) and guarantees that it vanishes for any $n$-qubit stabilizer state.
To complete the proof, we will now show that only pure stabilizer states can give zero GKP magic.

One can decompose arbitrary $n$-qubit states in the Hilbert-Schmidt basis as~\cite{altafini2004tensor}
\begin{align}
\label{eq:hilsch_dec}
    \hat{\rho}= \frac{1}{2^n}\qty(\hat{\mathds{1}} +\sum_{\hat{P}\in C_1^+ \setminus \mathds{1}} a_{\hat{P}} \hat{P})
\end{align}
with $a_{\hat{P}} \in \mathbb{R}$ and $\abs{a_{\hat{P}}} \leq 1$ while $C_1^+ \setminus \mathds{1}$ is the set of elements of the $n$-qubit Pauli group with only positive signs, without the identity. We only consider pure states, for which $\Tr\qty(\hat{\rho}^2)=1$. That is, $\sum_{P\in C_1^+ \setminus \mathds{1}}a_{\hat{P}}^2 = 2^n-1$, and one needs at least $2^n-1$ summands in the chosen decomposition to represent a pure state.

By definition, all pure states that have a stabilizer group $\mathcal{S} \subset C_1$ of cardinality $2^n$ are called stabilizer states, with $C_1$ the $n$-qubit Pauli group. That is, a stabilizer state $\hat{\sigma}$ is a $+1$ eigenstate of $2^n$ elements of the $n$-qubit Pauli group and can be written as
\begin{align}
    \hat{\sigma}= \frac{1}{2^n} \sum_{\hat{P} \in \mathcal{S}}\hat{P} = \frac{1}{2^n} \qty(\hat{\mathds{1}} +\sum_{\hat{P}\in \mathcal{S} \setminus \mathds{1}}\hat{P}).
\end{align}
Therefore, in the general $n$-qubit pure state decomposition of Eq.~(\ref{eq:hilsch_dec}), stabilizer states' coefficients satisfy $\abs{a_{\hat{P}}}=1$ for $\abs{\hat{P}} \in \mathcal{S}$, and zero otherwise. Therefore, pure stabilizer states have the minimal decomposition in the Hilbert-Schmidt basis.

The Wigner negativity per cell, related to the st-norm, for stabilizer states is
\begin{align}
    \cellneg_C(\hat{\sigma}) &= \frac{1}{\sqrt{\pi^n}}\sum_{\bm{i},\bm{j} \in \FF_2^n}\abs{ \Tr\qty[\hat{X}^{\bm{j}} \hat{Z}^{\bm{i}}   \hat{\sigma}]} = 
     \frac{1}{\sqrt{\pi^n}}\sum_{\bm{i},\bm{j} \in \FF_2^n} \abs{\frac{1}{2^n} \Tr\qty[ \hat{X}^{\bm{j}} \hat{Z}^{\bm{i}} \qty(\hat{\mathds{1}} +\sum_{\hat{P}\in \mathcal{S} \setminus \mathds{1}}\hat{P})]}\\ \nonumber
     &=\frac{1}{\sqrt{\pi^n}} \qty(1 + \sum_{\hat{P}\in \mathcal{S} \setminus \mathds{1}}  \abs{\frac{1}{2^n} \Tr\qty(\hat{ \mathds{1}} )})
     =\frac{1}{\sqrt{\pi^n}}(1+2^n-1) = \qty(\frac{2}{\sqrt{\pi}})^n.
\end{align}

An arbitrary pure non-stabilizer state $\hat{\rho}$ is given by
\begin{align}
\label{eq:hilsch_magic}
    \hat{\rho}= \frac{1}{2^n}\qty(\hat{\mathds{1}} +\sum_{\hat{P}\in \mathcal{M}^+ \setminus \mathds{1}}a_{\hat{P}} \hat{P})
\end{align}
where $\mathcal{M}\subseteq C_1^+$ is the set of Pauli operators with positive sign needed to represent the pure non-stabilizer state.
If we compare the cardinalities of the sets, it holds that $\text{card}\qty(\mathcal{S}) < \text{card}\qty(\mathcal{M})  \leq  \text{card}\qty(C_1^+)$, implying that the sum in Eq.~(\ref{eq:hilsch_magic}) involves more than $2^n$ terms. Indeed, we have shown that pure states need at least $2^n$ summands in the Hilbert-Schmidt decomposition, and that pure stabilizer states are expressed exactly with $2^n$ terms. Therefore, pure non-stabilizer states require necessarily more that $2^n$ summands. We notice that we cannot apply the same argument for the case of mixed states, since in general they can be expressed by less than $2^n$ terms.
The number of summands in the Hilbert-Schmidt representation of pure states is connected with the Wigner function of the corresponding GKP encoded states. We observe that higher number of summands correspond to higher number of Dirac-delta peaks in the Wigner function of the continuous-variable state~\cite{garcia-alvarez2021from}. That is, non-stabilizer states have a higher number of $w_{l,m}$ coefficients different from zero in Eq.~(\ref{eq:equallyspacedW}).

The Wigner negativity per cell for a general $n$-qubit non-stabilizer pure state $\hat{\rho}$ is
\begin{align}
\label{eq:wigneg_mag}
    \cellneg_C(\hat{\rho})=\frac{1}{\sqrt{\pi^n}} \sum_{\bm{i},\bm{j} \in \FF_2^n}\abs{ \Tr\qty( \hat{X}^{\bm{j}} \hat{Z}^{\bm{i}}   \hat{\rho})}&= \frac{1}{\sqrt{\pi^n}} \qty( 1 +  \sum_{\hat{P}\in \mathcal{M}^+ \setminus \mathds{1}} \abs{a_{\hat{P}}}).
\end{align}
We recall that pure states satisfy that $\sum_{\hat{P}\in \mathcal{M} \setminus \mathds{1}}a_{\hat{P}}^2 = 2^n-1$, and thus $a_{\hat{P}}^2 \leq 1$. Since  $\text{card}\qty(\mathcal{S}) <  \text{card}\qty(\mathcal{M})\leq \text{card}\qty(C_1^+)$, at least one coefficient in the Hilbert-Schmidt decomposition of pure non-stabilizer states of Eq.~(\ref{eq:hilsch_magic}) is strictly smaller than one, so that $a_{\hat{P}}^2 <1$ for at least one $\hat{P}$.  It holds that $\sqrt{x}>x$ for every positive real number $x$, such that  $x <1$. Therefore, in Eq.~(\ref{eq:wigneg_mag}) we have that
\begin{align}
    1 +  \sum_{\hat{P}\in \mathcal{M}^+ \setminus \mathds{1}} \abs{\sqrt{a_{\hat{P}}^2}} > 1 +  \sum_{\hat{P}\in \mathcal{M}^+ \setminus \mathds{1}} \abs{a_{\hat{P}}^2} = 1+2^n-1 = 2^n,
\end{align}
and the Wigner negativity per cell of a non-stabilizer state $\hat{\rho}$ is strictly larger than $\qty(\frac{2}{\sqrt{\pi}})^n$. The GKP magic defined in Eq.~(\ref{eq:GKPmana-second-expression}) is  consequently faithful.

\subsubsection{Invariance under composition with stabilizer state}
The property that the GKP magic is invariant under composition with stabilizer states follows directly from
\begin{align}
    \cellneg(\hat{\rho} \otimes \hat{\rho}_S)=\cellneg(\hat{\rho})
\end{align}
where we used faithfullness and multiplicativity.
This holds similarly for the GKP magic, where faithfullness and additivity is used.

\subsubsection{Non-increasing under computational basis measurement}
The final property missing is to show that the GKP magic is non-increasing under computational basis measurements or equivalently that
\begin{align}
    \sum_{\lambda} p_\lambda \cellneg_C(\rho_\lambda)\leq \cellneg_C(\hat{\rho}),
\end{align}
where $\lambda$ are measurement outcomes and with the following notation: The probability to measure outcome $\lambda$ is denoted by $p_\lambda = \Tr[\hat{\rho} \Pi_\lambda ]$ and the post-measurement state $\hat{\rho}_\lambda =\frac{1}{ p_\lambda}  \qty(\hat{\mathds{1}} \otimes \Pi_\lambda) \hat{\rho} \qty(\hat{\mathds{1}} \otimes \Pi_\lambda)      $ with $\Pi_\lambda$ being the POVM associated with outcome the measurement outcome.

Computational basis measurements in GKP are homodyne measurements. If the measured value is between $[2m\sqrt{\pi},(2mn+1)\sqrt{\pi})$ we are in logical 0 and if $[(2m+1)\sqrt{\pi}, 2m\sqrt{\pi})$ then in logical 1 for $m\in\mathcal{Z}$.
We define our POVMs for one cell and one mode as
\begin{align}
\begin{split}
    \Pi_0^{Cell} &= \int_0^{\sqrt{\pi}-\epsilon} dq \ket{q}\bra{q}\\
     \Pi_1^{Cell} &= \int_{\sqrt{\pi}}^{2\sqrt{\pi}-\epsilon} dq \ket{q}\bra{q}
     \end{split}
\end{align}
and thus for the entire space
\begin{align}
\begin{split}
    \Pi_0 &= \sum_m  \int_{2m\sqrt{\pi}}^{(2m+1)\sqrt{\pi}-\epsilon} dq \ket{q}\bra{q}\\
    \Pi_1 &= \sum_m  \int_{(2m+1)\sqrt{\pi}}^{2m\sqrt{\pi}-\epsilon} dq \ket{q}\bra{q}
    \end{split}
\end{align}
These POVM are both the identity in code space $\mathds{1}_{Code}$ as well as in the infinite dimensional Hilbertspace $\mathds{1}$. Both POVM elements are periodic with $2\sqrt{\pi}$ as well. The POVMs for more qubits can be decomposed in the ones defined above.

The absolute values of the Wigner function can be rewritten as
\begin{align}
\begin{split}
    &\abs{W_{  \sum_\lambda p_\lambda \hat{\rho}_\lambda } (\bm{r})}= \abs{\sum_\lambda p_\lambda W_{\hat{\rho}_\lambda}(\bm{r})}\\
    &=\abs{\sum_\lambda (2\pi)^{N_B} \int_{-\infty}^\infty \ddrn' W_{\hat{\rho}}(\bm{r}\oplus\bm{r}')W_{\Pi_\lambda}(\bm{r}' )     }\\
    &\leq  \sum_\lambda \abs{(2\pi)^{N_B} \int_{-\infty}^\infty \ddrn' W_{\hat{\rho}}(\bm{r}\oplus\bm{r}')W_{\Pi_\lambda}(\bm{r}' )     }\\
    &=   \sum_\lambda \abs{(2\pi)^{N_B} \lim_{n_1\rightarrow \infty} \cdots \lim_{n_{2N}\rightarrow \infty} n_1\dots n_{2N} \int_{\mathcal{C}'} \ddrn' W_{\hat{\rho}}(\bm{r}\oplus\bm{r}')W_{\Pi_\lambda}(\bm{r}' )     }\\
    &\leq \sum_\lambda (2\pi)^{N_B} \lim_{n_1\rightarrow \infty} \cdots \lim_{n_{2N}\rightarrow \infty} n_1\dots n_{2N} \int_{\mathcal{C}'} \ddrn' \abs{W_{\hat{\rho}}(\bm{r}\oplus\bm{r}')W_{\Pi_\lambda}(\bm{r}' )     } \\
    &= \lim_{n_1\rightarrow \infty} \cdots \lim_{n_{2N}\rightarrow \infty} n_1\dots n_{2N}  \int_{\mathcal{C}'} \ddrn' \abs{W_{\hat{\rho}}(\bm{r}\oplus\bm{r}') }
    \end{split}
\end{align}
for $N_B$ the dimension of the measured subsystem, $\mathcal{C}'$ the integral domain of the unit cell of the measured subsystem and
where we used that $\sum_\lambda \Pi_\lambda=\mathds{1}$ with $W[\mathds{1}](\bm{r})=\frac{1}{2\pi}^{N_B}$ and that the Wigner function of $\Pi_\lambda$ is translational invariant. We need to keep in mind that if we restrict to one unit cell, the measurement is only allowed to have support in one cell as well.
Thus we get
\begin{align}
\begin{split}
    \int_{-\infty}^\infty \ddrn \abs{\sum_\lambda p_\lambda W_{\hat{\rho}_\lambda}(\bm{r})} &=  \lim_{m_1\rightarrow \infty} \cdots \lim_{m_{2N}\rightarrow \infty} m_1\dots m_{2N} \int_{\mathcal{C}}\ddrn\abs{\sum_\lambda p_\lambda W_{\hat{\rho}_\lambda}(\bm{r})}\\
    &\leq \lim_{m_1\rightarrow \infty} \cdots \lim_{m_{2N}\rightarrow \infty} m_1\dots m_{2N} \int_{\mathcal{C}}\ddrn\sum_\lambda p_\lambda \abs{W_{\hat{\rho}_\lambda}(\bm{r})}\\
    &\leq \lim_{n_1\rightarrow \infty} \cdots \lim_{n_{2N}\rightarrow \infty} n_1\dots n_{2N}
    \lim_{m_1\rightarrow \infty} \cdots \lim_{m_{2N}\rightarrow \infty} m_1\dots m_{2N} \int_{\mathcal{C}} \ddrn \int_{\mathcal{C}'} \ddrn' \abs{W_{\hat{\rho}}(\bm{r},\bm{r}') }
    \end{split}
\end{align}
where we used that the post-measurements states are still valid code words.
Thus the Wigner negativity fulfills that
\begin{align}
\begin{split}
    \cellneg\qty[\sum_\lambda p_\lambda \hat{\rho}_\lambda] \leq \sum_\lambda p_\lambda  \cellneg\qty[\hat{\rho}_\lambda]  \leq \cellneg\qty[\hat{\rho}]\end{split}
\end{align}
Note again, by restricting the Wigner function to one unit cell, we get
\begin{align}
\begin{split}
    \cellneg_{C}\qty[\sum_\lambda p_\lambda \hat{\rho}_\lambda] \leq \sum_\lambda p_\lambda  \cellneg_{C}\qty[\hat{\rho}_\lambda]  \leq \cellneg_{C}\qty[\hat{\rho}].\end{split}
\end{align}
The GKP magic is therefore non-increasing under computational basis measurements as well, since the measurements in computational basis and post-selection conserve the purity of the input state. The post-measurement state will collapse to a stabilizer state in the measured subsystem.
So consequently with
\begin{align}
  \umag(\ket{\psi}) &=  \log_2 \qty(  \cellneg_C(\ket{\psi}\bra{\psi})) -n\log_2\qty(\frac{2}{\sqrt{\pi}})
\end{align}
we arrive at
\begin{align}
     \umag\qty(\sum_\lambda m_\lambda \ket{\psi}_\lambda) \leq \sum_\lambda m_\lambda  \umag\qty(\ket{\psi}_\lambda)  \leq \umag\qty(\ket{\psi})
\end{align}
where $m_\lambda$ is the probability amplitude associate with measurement outcome$\lambda$ and $\abs{m_\lambda}^2 =p_\lambda$.

\subsubsection{GKP magic and the discrete Wigner function}
\label{ap:discreteWigner}

The properties of the continuous-variable Wigner function differ from those of the discrete Wigner function making the former suitable for magic quantification for qubit systems (upon the mapping with GKP states that we introduced), unlike the latter. More specifically, in Ref.~\cite{veitch2014the-resource}, a magic measure based on the negativity of the discrete Wigner function, namely the sum negativity, is introduced for qudits, as well as for a single qubit. Even though it is possible to define a Wigner function for a multi-qubit system, its non-negativity is not preserved under all Clifford operations~\cite{raussendorf2017contextuality}. This property prohibits using the sum negativity as a magic measure for multi-qubit systems. In contrast to this, the properties of the CV Wigner function allow us to define a proper magic measure for multi-qubit systems upon the mapping based on the GKP encoding, satisfying, in particular, the required invariance under Clifford operations.
Here we explicitly explore the relation between the sum negativity of the discrete Wigner function and the GKP magic for a single qubit. 

The discrete Wigner of a single qubit and the Wigner function of a GKP code restricted to one unit cell consist of lattices with coefficients on the different sites.
So it is natural to compare the sum negativity of a discrete Wigner function to the GKP magic since both magic monotones are sums of elements on a lattice.

The discrete Wigner function for one qubit is defined as
\begin{align}
    \mathcal{C}_{\hat{\rho}}(\bm{u}) = \frac{1}{2} \Tr[\hat{A}_{\bm{u}} \hat{\rho}]
\end{align}
with 
\begin{align}
\begin{split}
    \hat{A}_{0}=\frac{1}{2}  \sum_{\bm{u}}  \hat{W}_{\bm{u}} \\
    \hat{A}_{\bm{u}}= \hat{W}_{\bm{u}} \hat{A}_0 \hat{W}_{\bm{u}}^\dagger
    \end{split}
\end{align}
where the discrete Wigner function can be thought of as a $2\times 2$ grid, $\bm{u} \in \mathcal{Z}_2\otimes \mathcal{Z}_2$ and 
 \begin{align*}
 \hat{W}_{l,m} = i^{l\cdot m} \hat{Z}^{m } \hat{X}^{l}
\end{align*}
are the Weyl operators.

The sum negativity is a magic monotone that is the sum of negative elements of the discrete Wigner function defined as
\begin{align}
    \text{sn}(\hat{\rho})=\frac{1}{2} \qty(\sum_{\bm{u}}\abs{ \mathcal{C}_{\hat{\rho}}(\bm{u}) }-1).
\end{align}

However, the Wigner negativity for a single qubit are given as
\begin{align}
\begin{split}
    \sum_{l,m=0}^1 \abs{w_{l,m}}= \sum_{l,m=0}^1 \abs{ \Vec{W}_{l,m}   \cdot \Vec{\rho}} &= \abs{\rho_{00} +\rho_{11}} + \abs{\rho_{00} -\rho_{11}} + \abs{\rho_{10} +\rho_{01}} + \abs{\rho_{01} -\rho_{10}} \\
    &= \abs{\Tr[\hat{\mathds{1}}\hat{\rho}]} + \abs{\Tr[ \hat{X} \hat{\rho} ]}+ \abs{\Tr[ \hat{Y} \hat{\rho} ]}+ \abs{\Tr[ \hat{Z} \hat{\rho} ]}.\end{split}
\end{align}

Hence, the sum negativity contains sums of Pauli operators with different relative signs. The Wigner negativity instead consists expectation values of a single Pauli operator.
\section{GKP magic and other magic quantifiers}
\label{ap:equivalent}
Finding an analytical expression for the GKP magic allows to go beyond what is possible by numerical brute force.
We rewrite the GKP magic in a more convenient form to derive analytical results. The Wigner negativity of one unit cell and thus the sums of the GKP magic can be rewritten as expectation values of Pauli operators
\begin{align}
\begin{split}
    &\sum_{\bm{i},\bm{j} \in \FF_2^n}\abs{\sum_{\bm{k} \in \FF_2^n}(-1)^{\bm{i}\cdot \bm{k}}  \rho_{\bm{k}, \bm{k}+\bm{j}}   }\\
     &= \sum_{\bm{i},\bm{j} \in \FF_2^n}\abs{\sum_{\bm{k},\bm{k'} \in \FF_2^n}(-1)^{\bm{i}\cdot \bm{k}} \rho_{\bm{k}, \bm{k'}}   \delta_{\bm{k'},\bm{k-j}}}\\
      &= \sum_{\bm{i},\bm{j} \in \FF_2^n}\abs{\sum_{\bm{k},\bm{k'} \in \FF_2^n}(-1)^{\bm{i}\cdot \bm{k}}  \rho_{\bm{k}, \bm{k'}}    \bra{\bm{k'}+\bm{j}}\ket{\bm{k}  }}\\
      &=\sum_{\bm{i},\bm{j} \in \FF_2^n}\abs{\sum_{\bm{k},\bm{k'} \in \FF_2^n}  (-1)^{\bm{i}\cdot \bm{k}}\rho_{\bm{k}, \bm{k'}}   \bra{\bm{k'}}\hat{X}^{\bm{j}} \ket{\bm{k}}  }\\
      &=  \sum_{\bm{i},\bm{j} \in \FF_2^n}\abs{\sum_{\bm{k},\bm{k'} \in \FF_2^n}\rho_{\bm{k}, \bm{k'}}    \bra{\bm{k'}}\hat{X}^{\bm{j}} \hat{Z}^{\bm{i}}   \ket{\bm{k}}  }\\
            &= \sum_{\bm{i},\bm{j} \in \FF_2^n}\abs{ \Tr\qty( \hat{X}^{\bm{j}} \hat{Z}^{\bm{i}}   \hat{\rho})}=2^n \; \mathcal{D}(\hat{\rho})
  \end{split}
\end{align}
where we used that subtraction is the same as addition in finite fields $\FF_2$ and $\mathcal{D}(\hat{\rho})$ is the st-norm~\cite{campbell2011catalysis}.

The st-norm was used to lower bound the robustness of magic~\cite{howard2017application}. We thus retrieved a known quantifier from magic state theory from a continuous-variable framework. We showed that it is related to the Wigner negativity of one unit cell via the proper normalization factor of $\sqrt{\pi}^n$. In particular, for pure states it holds the relation
\begin{align}
\umag(\ket{\psi}) = \log_2 \left( \mathcal{D}(\ket{\psi}\bra{\psi}) \right).
\end{align}
Given this form, we can connect the GKP magic as well with the stabiliser R\'enyi entropy~\cite{leone2022stabilizer}
\begin{align}
    M_{\alpha}(\ket{\psi}) = \qty(1-\alpha)^{-1}\log_2\qty(\sum_{\mathcal{P}\in C_1^+} 2^{-\alpha n} \bra{\psi}\mathcal{P}\ket{\psi}^{2\alpha}) -\log_2\qty(2^n),
\end{align}
where $\alpha\geq 0$ is a real parameter.
As the authors of Ref.~\cite{leone2022stabilizer} note, for $\alpha=1/2$ the stabilizer R\'enyi entropy is connected to the stabilizer norm
\begin{align}
     M_{\frac{1}{2}}(\ket{\psi})= 2 \log_2\qty(\mathcal{D}(\ket{\psi})),
\end{align}
which immediately gives the relation to the GKP magic
\begin{align}
    \umag(\ket{\psi}) = \frac{1}{2}M_{\frac{1}{2}}(\ket{\psi}).
\end{align}
Using this relation and the fact that $M_{\alpha}(\ket{\psi}) \leq \nu(\ket{\psi})$ with $\nu$ being the stabilizer nullity~\cite{beverland2020lower}, the GKP magic lower bounds the stabilizer nullity as
\begin{align}
     \umag(\ket{\psi})\leq \nu(\ket{\psi}).
\end{align}
Consequently the GKP magic lower bounds the robustness of magic as well as the stabilizer nullity.

\section{GKP magic and mixed states}
\label{ap:polytope}
The GKP magic defined before is only valid for pure states. In this section, we show that naively including mixed states into the definition does not work and give an explicit counter example. Furthermore, we derive the Wigner negativity for qubit states on the stabilizer polytope and the minimal value of the Wigner negativity.

We reduce Eq.~(\ref{eq_sum}) to the single qubit case and expand to
\begin{align}
    \sum_{l,m=0}^1 \abs{w_{l,m}}=  \abs{\rho_{00} +\rho_{11}} + \abs{\rho_{00} -\rho_{11}} + \abs{\rho_{10} +\rho_{01}} + \abs{\rho_{01} -\rho_{10}} ,
\end{align}
where the first summand is obviously just the trace.
The faces of the stabilizer polytope have to fulfill the condition
\begin{align}
    \hat{\rho}_P = x \ket{X}\bra{X} +y \ket{Y}\bra{Y} +z \ket{Z}\bra{Z}
\end{align}
with $x,y,z \in [0,1]$, $x+y+z =1$ and $\ket{X/Y/Z}$ being either one of the eigenstates of the corresponding Pauli operator $\hat{X}, \hat{Y}, \hat{Z}$.
Plugging this state in the sum, we get
\begin{align}
\sum_{l,m=0}^1  \abs{w_{l,m}} &= 1+ \abs{\rho_{00} -\rho_{11}} + \abs{\rho_{10}+ \rho_{01}} + \abs{\rho_{01} -\rho_{10}} \\   
 &= 1+z +x+y =2.
\end{align}
Thus, if we subtract $\cellnegint=\frac{2}{\sqrt{\pi}}$ per qubit, the stabilizer polytope has 0 GKP magic. 

Using this analysis, it is easy to see that the  Wigner logarithmic  negativity of one cell for the maximally mixed state $\hat{\rho}_M = \frac{1}{2} \qty(\ket{X/Y/Z}\bra{X/Y/Z} + \ket{-X/Y/Z}\bra{-X/Y/Z}   )  $ corresponds to  
\begin{align}
 \sum_{l,m=0}^1  \abs{w_{l,m}} = 1  , 
\end{align}
which is the lowest value possible.

So there exist states with different negativity that should map to a single value for a proper magic measure. This can be done by setting them manually to the same value, e.g.
\begin{align}
        \Tilde{\umag}(\hat{\rho}) = \max\qty[0, \log_2\qty(\frac{1}{\sqrt{\pi}^{n}}  \sum_{\bm{i},\bm{j} \in \FF_2^n}\abs{\sum_{\bm{k} \in \FF_2^n}(-1)^{\bm{i}\cdot \bm{k}}  \rho_{\bm{k}, \bm{k}+\bm{j}}   } ) - \log_2 \qty[\cellnegint(n)]].
\end{align}

Some mixed states may have lower values than pure stabilizer states, as we have seen for one qubit. 
Let us assume a state $\hat{\rho}= \hat{\rho}_I \otimes \hat{\rho}_O$ with $\hat{\rho }_I$ being inside the polytope and thus having lower cell negativity than a stabilizer state and  $\hat{\rho }_O$ being outside the polytope and thus having larger negativity than a pure stabilizer state.
We furthermore assume that the difference between the negativity of a pure stabilizer state and  $\hat{\rho }_I$ is larger than the one between a pure stabilizer state and $\hat{\rho }_O$. An example is the product between the maximally mixed state $\hat{\rho}_0 = \frac{1}{2}\qty(\ket{0}\bra{0}+\ket{1}\bra{1})$ and the $\ket{H}$ state for a state outside the polytope. Then we get
\begin{align}
\begin{split}
    \Tilde{\umag}(\hat{\rho}_I \otimes \hat{\rho}_O) &= \max\qty[0, \log_2\qty( \int_{\mathcal{C}_1} \ddrn_1\abs{ W_{\hat{\rho}_I} } \int_{\mathcal{C}_2} \ddrn_2 \abs{ W_{\hat{\rho}_O} }   ) - \log_2 \qty[\cellnegint(n)]]\\
    &=  \max\qty[0, \log_2\qty( \int_{\mathcal{C}_1} \ddrn_1\abs{ W_{\hat{\rho}_I} } )+ \log_2\qty(\int_{\mathcal{C}_2} \ddrn_2 \abs{ W_{\hat{\rho}_O} } )   - \log_2 \qty[\cellnegint(n)]]    \\
    &=\max\qty[0, -1 +0.272]\\
    &= 0.\end{split}
\end{align}
The last equality arises because the second argument is negative. 
A magic measure needs to be invariant under composition with a stabilizer state and therefore should just give the resource content of the $H$ state.

\subsubsection{Wigner negativity and partial trace}
It is possible to show that the Wigner negativity is non-increasing under a partial trace operation. Since the GKP magic cannot be defined for mixed states, the partial trace is excluded from the list of CKP Magic non-increasing operations.
It holds for the Wigner function that
\begin{align}
    W_{\Tr_2[ \hat{\rho}_{12} ]} (\bm{r}_1)= \int_{-\infty}^\infty \ddrn_2  W_{\hat{\rho}_{12} } (\bm{r}_1, \bm{r}_2).
\end{align}
So the Wigner negativity is given as
\begin{align}
\begin{split}
    &\int_{-\infty}^\infty \ddrn_1 \abs{W_{\Tr_2[ \hat{\rho}_{12} ]} (\bm{r}_1)} = \int_{-\infty}^\infty \ddrn_1 \abs{\int_{-\infty}^\infty \ddrn_2  W_{ \hat{\rho}_{12} } (\bm{r}_1, \bm{r}_2)}\\
    &= \lim_{n_1\rightarrow \infty} \cdots \lim_{n_{2N}\rightarrow \infty} n_1\dots n_{2N}
    \lim_{m_1\rightarrow \infty} \cdots \lim_{m_{2N}\rightarrow \infty} m_1\dots m_{2N} \;\int_{\mathcal{C}_1} \ddrn_1 \abs{\int_{\mathcal{C}_2} \ddrn_2  W_{ \hat{\rho}_{12} } (\bm{r}_1, \bm{r}_2)}\\
    &\leq \lim_{n_1\rightarrow \infty} \cdots \lim_{n_{2N}\rightarrow \infty} n_1\dots n_{2N}
    \lim_{m_1\rightarrow \infty} \cdots \lim_{m_{2N}\rightarrow \infty} m_1\dots m_{2N} \;\int_{\mathcal{C}_1} \ddrn_1 \int_{\mathcal{C}_2}\ddrn_2  \abs{ W_{\hat{\rho}_{12} } (\bm{r}_1, \bm{r}_2)}\\
    &=\lim_{n_1\rightarrow \infty} \cdots \lim_{n_{2N}\rightarrow \infty} n_1\dots n_{2N}
    \lim_{m_1\rightarrow \infty} \cdots \lim_{m_{2N}\rightarrow \infty} m_1\dots m_{2N} \;\cellneg_{C}(\hat{\rho}_{12})
    \end{split}
\end{align}
where we used the integral triangle equality.

The last step missing is to show how the partial trace acts on our code space and thus on one unit cell.
The partial trace on the code space is
\begin{align}
\begin{split}
    \Tr_2[\hat{\rho}_{12}] &= \sum_{\bm{i}\in \FF_2^n} \bra{\bm{i}}_2 \hat{\rho}_{12}\ket{\bm{i}}_2\\
  &= \int_{-\infty}^\infty \ddqn \bra{q}_2 \hat{\rho}_{12}\ket{q}_2\\
  &= \lim_{n_1\rightarrow \infty} \cdots \lim_{n_{2N}\rightarrow \infty} n_1\dots n_{2N} \int_{\mathcal{C}} \ddqn \bra{q}_2 \hat{\rho}_{12}\ket{q}_2\\
  &= \lim_{n_1\rightarrow \infty} \cdots \lim_{n_{2N}\rightarrow \infty} n_1\dots n_{2N}  \Tr_2[\hat{\rho}_{12}]_{\mathcal{C}}.
  \end{split}
\end{align}
Hence, we need to keep that in mind that for consistency the trace should only act within the unit cell if we restrict it to a unit cell.

Consequently, the Wigner negativity of one cell is non-increasing under the action of a partial trace
\begin{align}
    \cellneg_{C}(\Tr_2[\hat{\rho}_{12}]) \leq\cellneg_{C}\qty(\hat{\rho}_{12}).
\end{align}

\subsubsection{Lower bounds for probabilistic gate synthesis}

As described in the main text, we can lower bound probabilistic stabilizer protocols that convert $k$ copies of an $r$-qubit state $\ket{\psi}$ to $m$ copies of the $s$-qubit target state $\ket{\phi}$ with probability $p$.
The following calculations will clarify the connection between the st-norm~\cite{campbell2011catalysis} and the GKP magic, namely that the logarithm of the st-norm ---the Wigner logarithmic negativity restricted to one GKP cell, $\logneg_C$--- lower bounds the GKP magic. Notice that we can define the GKP magic for mixed states mathematically, although it is not a magic measure. We highlight the parallelism with this lower bound and the lower bounds on the Robustness of Magic with the st-norm~\cite{howard2017application}.

In Sec.~\ref{ap:properties} and~\ref{ap:polytope}, we show that the Wigner logarithmic negativity of one unit cell is additive and does not increase under probabilistic stabilizer protocols, so that 
\begin{align}
    k\logneg_{C}(\ket{\psi}) \geq  p m \logneg_{C}(\ket{\phi}).
\end{align}
For the average number of copies $\EE[n]$ of $\ket{\psi}$ needed to distill $\ket{\phi}^{\otimes m}$ it has to hold that
\begin{align}
    \EE[n] = \frac{k}{p} \geq m \frac{\logneg_{C}(\ket{\phi})}{\logneg_{C}(\ket{\psi}) },
\end{align}
where the protocol has to be run $1/p$ times for a successful outcome.

The Wigner logarithmic negativity per cell is directly related with the GKP magic as $\umag(\ket{\Psi})=\logneg_C(\ket{\Psi})-\log_2[\cellnegint(n)]$, where we have subtracted the intrinsic logarithmic negativity per cell of a pure $n$-qubit stabilizer state, given by $\log_2[\cellnegint(n)]=(2/\sqrt{\pi})^n$. Thus, the bound can be rewritten as
\begin{align}
    \EE[n] = \frac{k}{p} \geq m \frac{\umag(\ket{\phi})+\log_2[\cellnegint(s)] }{\umag(\ket{\psi})+\log_2[\cellnegint(r)] }.
\end{align}
For a deterministic protocol ($p=1$), this bound simplifies to
\begin{align}
\label{eq:boundWlog}
    \frac{k}{m}\geq \frac{\umag(\ket{\phi})}{\umag(\ket{\psi})}\qty[\log_2[\cellnegint(s)]-\frac{k}{m} \log_2[\cellnegint(r)] ].
\end{align}
On the other hand, the GKP magic bound for deterministic protocols is
\begin{align}
\label{eq:boundGKPmag}
    \frac{k}{m} \geq  \frac{\umag(\ket{\phi}) }{\umag(\ket{\psi})}.
\end{align}

We can thus compare both bounds of Eqs.~(\ref{eq:boundWlog}) and (\ref{eq:boundGKPmag}) to determine which one is tighter. In particular, the lower bound given by the GKP magic in Eq.~(\ref{eq:boundGKPmag}) is higher when
\begin{align}
     1 \geq \qty[\log_2[\cellnegint(s)]-\frac{k}{m} \log_2[\cellnegint(r)]] =  \log_2\qty[\frac{\cellnegint(s)}{\cellnegint(r)^{k/m}}] = \log_2\qty[\qty(2/\sqrt{\pi})^{s-rk/m}].
\end{align}
Namely, when
\begin{align}
\label{eq:conditionbound}
    s-r\frac{k}{m} &\leq \qty[\log_2(2/\sqrt{\pi})]^{-1}.
\end{align}
We observe that the inequality in Eq.~(\ref{eq:conditionbound}) always holds since the number of initial qubits $kr$ is always higher or equal to the number of output qubits $ms$. Thus, $kr/m \geq s$ and $s - kr/m \leq 0 < \qty[\log_2(2/\sqrt{\pi})]^{-1}$.

 Hence, the lower bound provided by the Wigner logarithmic negativity per cell is strictly lower (less tight) than the lower bound given by the GKP magic.

\section{Lower bounds and GKP magic of computational tasks}

\subsubsection{GKP magic of the Quantum Adder and Quantum Fourier Transform}
\label{ap:adder}
The Quantum Adder and the Quantum Fourier Transform are building blocks for many quantum algorithms of practical interest. Note that we will retrieve the analytical value of the GKP magic for a $H$ state, the state that teleports a $T$-gate for free during the derivation.
Using the techniques described in the main text about distillation and gate synthesis, we can lower bound the $T$-count of these building blocks.
The derivation of analytical values of the GKP magic for a quantum adder follows parallel work in Ref. \cite{beverland2020lower} closely.
The modular adder is a fundamental building block e.g. in Shor's algorithm. The adder circuits act on two qubit registers as
\begin{align}
\label{eq:adder}
    A(\ket{\bm{i}}\ket{\bm{j}}) = \ket{\bm{i}}\ket{\bm{i}+\bm{j}}
\end{align}
where both states are $n$-qubit registers and the addition is intended upon mod $2^n$. 

Another ingredient are the family of quantum Fourier states for integers $a$
\begin{align}
    \ket{QFT^a_n}=\bigotimes^n_{k=1} \frac{\ket{0}+e^{i2\pi a/ 2^k}\ket{1}}{\sqrt{2}}.
\end{align}
The QFT states show the action of the adder $A$ on a stabilizer state $\ket{+}^{\otimes n}$ and a $n$-qubit QFT state $\ket{QFT^b_n}$
\begin{align}
    A(\ket{+}^{\otimes n} \ket{QFT^b_n}) = \ket{QFT^{-b}_n} \ket{QFT^b_n}.
\end{align}
This is straightforward to show by calculating $A(\ket{QFT^a_n} \ket{QFT^b_n})$ and then setting $a=0$. The calculation can be found in~\cite{beverland2020lower}.
Hence, to bound the resources needed to implement the adder, it needs to hold that for some input resource state $\ket{\psi}$
\begin{align}
    \umag(\ket{\psi} \ket{QFT^b_n} ) \geq \umag(\ket{QFT^{-b}_n}\ket{QFT^b_n})
\end{align}
and by using additivity we obtain
\begin{align}
    \umag(\ket{\psi} ) \geq \umag(\ket{QFT^{-b}_n}).
\end{align}
By setting $b=1$ for simplicity, we find a bound for the adder
\begin{align}
    \umag(\ket{QFT^{-1}_n})
\end{align}
and the quantum Fourier transform as well.

The GKP magic for the states $\ket{QFT^{-1}_n}$ is given as
\begin{align}
\begin{split}
  \umag(\ket{QFT^{-1}_n}) &= \umag \qty( \bigotimes^n_{k=1} \frac{\ket{0}+e^{-i2\pi/ 2^k}\ket{1}}{\sqrt{2}} ) \\
  &=\sum_{k=1}^n \umag \qty( \frac{\ket{0}+e^{-i2\pi / 2^k}\ket{1}}{\sqrt{2}} ),\end{split}
\end{align}
where we used the additivity of the GKP magic.
So we just need to calculate
\begin{align}
    \umag \qty( \frac{\ket{0}+e^{-i2\pi / 2^k}\ket{1}}{\sqrt{2}} )
\end{align}
which are just single qubit states.
In order to derive an analytical value, we insert this single qubit state into the alternative form of the Wigner negativity defined in Sec.~\ref{ap:equivalent}.
Consequently, we obtain for the GKP magic
\begin{align}
     \umag\qty(\ket{QFT^{-1}_n}) = \sum_{k=1}^n \log_2\qty( \frac{1}{\sqrt{\pi}}   \qty(1+ \qty(\abs{\sin\qty(\frac{2\pi}{2^k})}+ \abs{\cos\qty(\frac{2\pi}{2^k})}    )))  - n\log_2 \qty(\frac{2}{\sqrt{\pi}}).
\end{align}
\newcommand\myeq{\mathrel{\stackrel{\sim}{\makebox[0pt]{\mbox{\normalfont\scriptsize n $\to \infty$}}}}}

\begin{figure}
    \centering
    \includegraphics{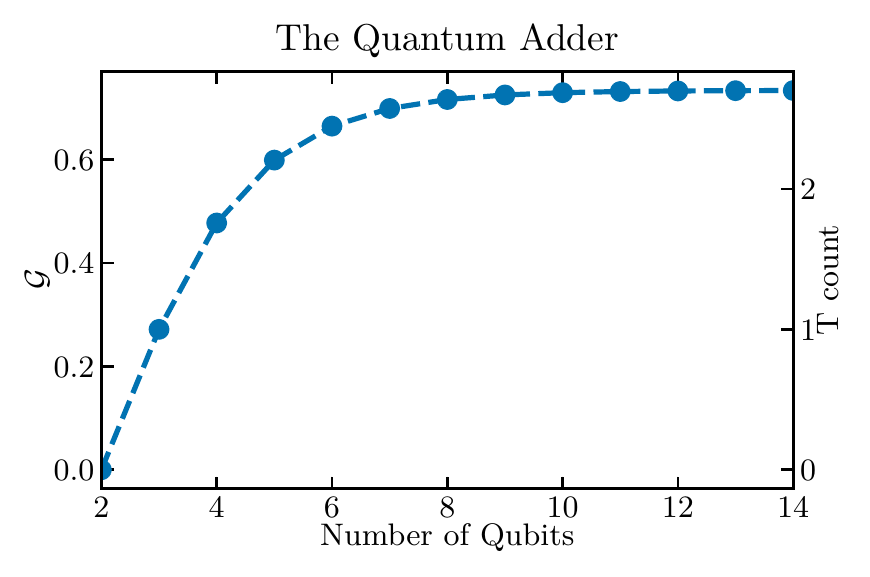}
    \caption{The GKP magic for various numbers of qubits for the quantum adder defined in Eq.~\ref{eq:adder}. We use the resource conent of the $\ket{H}$ state to lower bound the $T$-gates needed to implement the quantum adder. It can be seen that for our measure the resource content needed to implement the adder approaches a finite value.}
    \label{fig:adder}
\end{figure}

We can analytically bound the monotone from above
\begin{align}
     \umag\qty(\ket{QFT^{-1}_n})\leq  n \log_2\qty(\frac{1+\sqrt{2}}{\sqrt{\pi}}) - n\log_2 \qty(\frac{2}{\sqrt{\pi}})
\end{align}
where $\sqrt{2}$ is the maximum of $\abs{\sin\qty(\frac{2\pi}{2^k})}+ \abs{\cos\qty(\frac{2\pi}{2^k})} $. 
Note that this value corresponds to the value of one $H$ state $\umag(\ket{H} = \log_2\qty(\frac{1+\sqrt{2}}{2})$.
The exact values are shown in Fig.~\ref{fig:adder}.

For large values of the integer $k$, the summands go asymptotically to one
\begin{align}
    \abs{\sin\qty(\frac{2\pi}{2^k})}+ \abs{\cos\qty(\frac{2\pi}{2^k})}   \rightarrow 1
\end{align}
which implies the the measure will converge to a finite number as can be seen in Fig.~\ref{fig:adder}.

\subsubsection{Multiply-controlled phase gate}
\label{ap:phase gate}
Most gate sets contain controlled gates as their required multiqubit gates. Especially the diagonal gates from the third level of the Clifford hierarchy are of interest because they allow for easy implementation through teleportation gadgets. A family of interest for these gates is the multiply controlled phase gates. In this section, we derive the analytical value of the GKP magic for this type of gate.

The multiply-controlled phase gate is represented in computational basis as 
\begin{align}
  M_\phi=\text{diag}(1,\dots,1,e^{i \phi}).
\end{align}
Since this family of unitaries are diagonal and belong to the third level of the Clifford hierarchy, we can use the same strategy as before and calculate
\begin{align}
    \ket{ M_\phi} =  \hat{M}_\phi \ket{+}^{\otimes n}= \frac{1}{\sqrt{2^n}}\sum_{\bm{x} \in \FF_2^n} \qty[e^{i \phi}]^{x_1\cdots x_n}\ket{\bm{x}}.
\end{align}
We can calculate the GKP magic analytically for these states by using the alternative form of the Wigner negativity defined in Sec.~\ref{ap:equivalent}
\begin{align}
\begin{split}
    &2^n\cdot \bra{ M_\phi}\hat{X}^{\bm{i}} \hat{Z}^{\bm{j}}\ket{ M_\phi} \\ 
    &= \sum_{\bm{x},\bm{x'}} \qty[e^{i \phi}]^{x_1\cdots x_n}  \qty[e^{-i \phi}]^{x'_1\cdots x'_n} \bra{x'} X^{\bm{i}} Z^{\bm{j}}\ket{\bm{x}}\\
    &=   \sum_{\bm{x}} \qty[e^{i \phi}]^{x_1\cdots x_n}  \qty[e^{-i \phi}]^{(x_1+i_1)\cdots (x_n+i_n)} (-1)^{\bm{j}\cdot \bm{x}} \\
    &= \sum_{v} \qty[e^{i \phi}]^{x_1\cdots x_n-(x_1+i_1)\cdots (x_n+i_n)}  (-1)^{\bm{j}\cdot \bm{x}}.\end{split}
\end{align}
For $\bm{i}=O^n$, the only term that contributes is $\sum_{\bm{x}} (-1)^{\bm{j}\cdot \bm{x}}$, which is $2^n$ for $\bm{j}=0^n$ and 0 for the rest.
For  $\bm{i}\neq 0^n $, only for $\bm{x}=1^n$ and $\bm{x}=1^n+\bm{i}$, the summation is different from $\sum_{\bm{x}} (-1)^{\bm{j}\cdot \bm{x}}$. So we have
\begin{align}
\begin{split}
  &2^n\cdot   \sum_{\bm{x}} \qty[e^{i \phi}]^{x_1\cdots x_n-(x_1+i_1)\cdots (x_n+i_n)}  (-1)^{\bm{j}\cdot \bm{x}}\\
  &=  e^{i\phi} (-1)^{\bm{j}\cdot 1^n} +e^{-i\phi} (-1)^{\bm{j}\cdot (1^n+\bm{i})}+\sum_{x\setminus\{1^n,1^n+\bm{i} \}} (-1)^{\bm{j}\cdot \bm{x}}\\
  &=  (e^{i\phi}-1) (-1)^{\bm{j}\cdot 1^n} +(e^{-i\phi}-1) (-1)^{\bm{j}\cdot (1^n+\bm{i})} +\sum_{x} (-1)^{\bm{j}\cdot \bm{x}}.\end{split}
\end{align}
Thus for $\bm{j}=0^n$, this expression is $ 2^n+ (e^{i\phi}-1) +(e^{-i\phi}-1)$. Otherwise for $\bm{j}\cdot \bm{i}$ even, the sum is $ (-1)^{\bm{j}\cdot 1^n}   [  ( (e^{i\phi}-1) +(e^{-i\phi}-1) )]$ and for $\bm{j}\cdot \bm{i}$ odd, the sum is $ (-1)^{\bm{j}\cdot 1^n}   [  ( (e^{i\phi}-1) -(e^{-i\phi}-1) )]$.

So to summarize 
\begin{align}
\begin{split}
    \abs{\bra{ M_\phi} \hat{X}^{\bm{i}} \hat{Z}^{\bm{j}} \ket{ M_\phi}} = \begin{cases}
    1 & \text{if $\bm{i}=0^n$ and $\bm{j}=0^n$}\\
     0 & \text{if $\bm{i}=0^n$ and $\bm{j}\neq0^n$ }\\
      \abs{1+2^{-n}(e^{i\phi}-1) +(e^{-i\phi}-1) }   & \text{if $\bm{i}\neq 0^n$ and $\bm{j}=0^n$}\\
      2^{-n}\abs{(2\cos(\phi)-2 )} & \text{if $\bm{i}\neq0^n$ and $\bm{j}\neq 0^n$ and $\bm{i}\cdot \bm{j}$ even}\\
        2^{-n}\abs{(2\sin(\phi) )} & \text{if $\bm{i}\neq0^n$ and $\bm{j}\neq 0^n$ and $\bm{i}\cdot \bm{j}$ odd}\\
    \end{cases}\end{split}
\end{align}
The mulitplicity for the first case is 1, for the second $2^n-1$, for the third $2^n-1$, for the fourth $1-3\cdot 2^{n-1} +2^{2n-1}$ and for the fifth  $2^{2n-1}-2^{n-1}$.
As can be seen, the GKP magic goes asymptotically to a finite value for increasing numbers of qubits, since the contribution of the additional qubits goes asymptotically to 0.

\subsubsection{GKP magic using the Choi–Jamiołkowski isomorphism}
\label{ap:choi}
A different route to quantify the resource of unitary operations than using the teleportation circuit for unitaries which are diagonal from the third level of the Clifford hierarchy, is to employ the Choi–Jamiołkowski isomorphism~\cite{choi1975completely, jamiokowski1972linear}. We use this to calculate the resource content of gates that are not diagonal unitaries from the third level of the Clifford hierarchy.

The maximally entangled state for two qubits is given by
\begin{align}
    \ket{\psi}=\frac{1}{\sqrt{2}}\sum_{i=0}^{1} \ket{i,i}.
\end{align}
For a general completely positive-map $\Phi$, the Choi state is given by
\begin{align}
    \hat{\varphi}= (\Phi \otimes \mathds{1}) \ket{\psi} \bra{\psi} = \frac{1}{2}\sum_{j,k=0}^{1} \Phi (\ket{j}\bra{k}) \otimes \ket{j}\bra{k}.
\end{align}
The map can be retrieved probabilistically by a teleportation procedure. For our purposes, however this is not important. For a trace-preserving map $\hat{\Phi}$ the Choi state $ \varphi$ is a proper normalised state, if $\Phi$ is unitary then  $\hat{\varphi}$ is a pure quantum state.
Thus if we define $\Phi/qty(\hat{\rho}) = \hat{U}  \hat{\rho} \hat{U}^\dagger$ with $\hat{U} \in \text{SU}(2^n)$, we can calculate the GKP magic for arbitrary unitaries.
For a $n$-qubit unitary, the Choi state is then given by
\begin{align}
    \ket{\varphi_U}= (\hat{U}\otimes \mathds{1}) \frac{1}{\sqrt{2^n}} \sum_{\bm{j} \in \FF_2^n}\ket{\bm{j},\bm{j}}.
\end{align}

\subsubsection{GKP magic maximization of general states and unitaries}
\label{ap:MostMagic}

\begin{figure}
    \centering
    \includegraphics{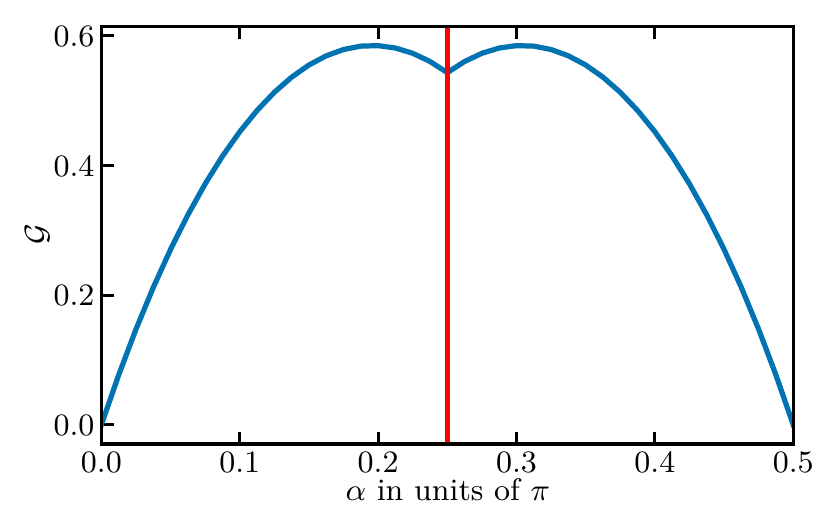}
    \caption{The GKP magic for various angle $\alpha$ of the parametrization defined in Eq.~(\ref{eq:1q_par}) and fixed angles $\phi_1=0$ and $\phi_2 = \frac{\pi}{4}$. We find two maximas in the investigated interval that are at the same time the maximal value for all possible single qubit unitaries. The red line symbolizes the angle for which the robustness of magic is maximized~\cite{howard2017application}. As can be seen, our measure is in a local minimum for this angle.}
    \label{fig:mostmagic}
\end{figure}

Instead of asking what the resource content of a particular state is, we can ask what is the most magic state.
We can find the most magic state by numerically optimizing over all states of constant size and maximize the GKP magic.

The most magic single qubit state with $\umag(\ket{\Psi_1})=0.450$ corresponds to the state $\ket{T}$
\begin{align*}
    \ket{\Psi_1}=\ket{T}= \cos(\beta) \ket{0} + \sin(\beta) e^{i\frac{\pi}{4}} \ket{1}, \; \cos(2\beta)=\frac{1}{\sqrt{3}}.
\end{align*}

For two qubits, the most magic state is given as
\begin{align}
\ket{\Psi_2}=\begin{pmatrix}
      -0.2839+i\cdot0.2933\\
    -0.01299-i\cdot 0.7886\\
    0.003388+i\cdot0.2112\\
    -0.2936-i\cdot 0.2838 
    \end{pmatrix}
\end{align}
 with the GKP magic being $\umag(\ket{\Psi_2})= 0.900$.

For three qubits, the most magic state is the Hoggar state  
\begin{align}
    \ket{\Psi _3}=\ket{\text{Hoggar}} = \frac{1}{\sqrt{6}}\begin{pmatrix}
    1+i\\
    0\\
    -1\\
    1\\
    -i\\
    1\\
    0\\
    0
    \end{pmatrix}
\end{align}
with  $\umag(\ket{\Psi_3})= 1.459$.

Using the Choi–Jamiołkowski isomorphism described in Sec.~\ref{ap:choi}, we can calculate the most magic unitary as well.
Since we are mapping a unitary into states involving twice the number of qubits on which the unitary has support, the calculated values need to be smaller or equal than the most magic state of this number of qubits.

A general $2\times 2$ unitary matrix can be parametrized as
\begin{align}
\label{eq:1q_par}
    U_2 = \begin{pmatrix}
    e^{i \phi_1} \cos(\alpha) &    e^{i\phi_2} \sin(\alpha) \\
    -e^{-i\phi_2} \sin(\alpha) & e^{-i \phi_1} \cos(\alpha) 
    \end{pmatrix}.
\end{align}
Thus we choose to represent a single qubit unitary operation $\hat{U} \in \text{SU}(2) $ in the computational basis in the above-defined parametrization.
By numerically optimizing over all one qubit unitaries, we found the maximal GKP magic with  $\umag(\ket{\varphi_{U^m}}= 0.585$ at the optimal angles $\phi_1=0 ,\phi_2=\frac{\pi}{4} , \alpha=0.6155$  $\frac{\pi}{6}<\alpha < \frac{\pi}{5}$. The angles agree well with~\cite{howard2017application} except that we do not have $\alpha = \frac{\pi}{4}$. A plot showing this difference can be seen in Fig.~\ref{fig:mostmagic}.

In order to characterize the most magic multi-qubit unitary, we parametrize the unitaries by using the Cartan decomposition of $\text{SU}(2^n)$~\cite{khaneja2000cartan} with $n$ being the number of qubits. A explicit expression for the two qubit $U_4 \in \text{SU}(2^2) $ case is given in~\cite{vatan2004optimal}.

The most GKP magic two-qubit unitary is then given by
\begin{align}
\ket{\phi_{U^m}}=
\begin{pmatrix}
    0.2596-i \cdot0.01.088   & 0.09603-i\cdot 0.5987  &  0.1001-i\cdot 0.2.452    & 0.08229-i\cdot0.6985\\
    0.3256-i\cdot 0.2647   & -0.3450e- i\cdot0.1906   & -0.2223- i\cdot 0.4.320    &0.5056+ i\cdot0.4205\\
    -0.2656- i\cdot 0.4279    & 0.4459-i \cdot0.5166    & -0.3684+i\cdot0.2692    & -0.08724+i\cdot 0.2546\\
    -0.1566-i \cdot0.6915    &-0.005356+i\cdot0.1057    & 0.6923-i \cdot0.07978    & 0.01901-i \cdot0.009329
    \end{pmatrix}
\end{align}
with $\umag(\ket{\varphi_{U^m}}=1.728$.

\subsubsection{GKP magic comparison and numerical results}
\label{ap:num}

\onecolumngrid
In this section, we collect numerical values calculated with the GKP magic. A list of quantum states with their resource content can be found in Table~\ref{tab:num_res}.
Additionally to the gates described in the main text, whose $T$-count has been calculated in~\cite{mosca2021a-polynomial}, we also compute the resource of the following 4-qubit circuits
\begin{align}
    \hat{U}_1 &= (CCX \otimes \hat{\mathds{1}}) (\hat{\mathds{1}}\otimes CCX)\nonumber \\
    \hat{U}_2 &= (CCX \otimes \hat{\mathds{1}}) (\hat{\mathds{1}}\otimes CCX)(CCX \otimes \hat{\mathds{1}}).
\end{align}
The results are given in Table~\ref{tab:choi_ap}. The $T$-counts are again lower than the ones provided in~\cite{mosca2021a-polynomial}, but those are given for unitary synthesis.
However, we reproduce that $U_1$ has more resource than $U_2$ thus the same hierarchy as in~\cite{mosca2021a-polynomial}.

\twocolumngrid

\begin{table}
\caption {Comparison between the  GKP magic and the Robustness of Magic (RoM)~\cite{howard2017application} for states of the form $\ket{U} = \hat{U} \ket{+}^{\otimes n}$. The unitaries $\hat{U}$ are diagonal gates from $C_3$ and admit a resourceless implementation using the states $\ket{U}$. The GKP-Magic and the RoM give the same $T$-count.} \label{tab:num_res}

\begin{ruledtabular}
\begin{tabular}{cccc}
 $\hat{U}$ & Robustness of Magic &  GKP magic &$T$-count \\ \hline
$T_1$ &1.41421& 0.272 & 1\\
$T_{1,2}$&1.74755& 0.543&2\\
$CS_{12}$&2.2&0.807  & 3 (2.967)\\
$T_{1,2,3}$&2.21895&0.815& 3 (2.996)\\
$CS_{12,13}$&2.55556&0.907&4 (0.907)\\
$T_1 CS_{23}$&2.80061&1.079& 4 (3.966)\\
$T_1 CS_{12,13}$&3.12132&1.195&  5 (4.393)\\
$C^2Z$& &0.907& 4 (3.335)\\
$C^3Z$& & 1.267& 5 (4.658)\\
$C^4Z$& &1.431 &6 (5.261)\\
$C^2S$& &1.210& 5 (4.449)\\
$C^3S$& & 1.401& 6 (5.151)\\
$C^4S$& & 1.494& 6 (5.493)\\
\end{tabular}
\end{ruledtabular}
\end{table}
\begin{table}
\caption {GKP magic computed using the Choi–Jamiołkowski isomorphism for gates analyzed classically with unitary synthesis protocols~\cite{mosca2021a-polynomial}. In general, a unitary involving $n$ qubits is mapped to a $2n$-qubit state.
The largest system size we consider involves $12$ qubits and corresponds to the $C^{5}X$ gate.
We do not recover the same $T$-cost given in~\cite{mosca2021a-polynomial}, since the GKP magic admits more general gate synthesis than unitary synthesis. The same hierarchy however is given. } \label{tab:choi_ap}
\begin{ruledtabular}
\begin{tabular}{crr}
 $U$ &  GKP magic & $T$-count\\ \hline
 Toffoli& 0.907 & 4 (3.335)\\
Fredkin & 0.907&4 (3.335)\\
$C^3X$& 1.2667& 5 (4.657)\\
$C^4X$& 1.431& 6 (5.261)\\
$C^5X$& 1.509& 6 (5.548)\\
$U_1$&1.570&  6 (5.772)\\
$U_2$ &0.907& 4 (3.335)\\
\end{tabular}
\end{ruledtabular}
\end{table}


\end{document}